\documentclass[12pt]{article}
\usepackage{a4wide}
\usepackage[centertags]{amsmath}
\usepackage{amssymb}
\usepackage[sort&compress,numbers]{natbib}
\usepackage{ifpdf}
\usepackage{setspace}
\usepackage{amsfonts}
\usepackage{color}
\usepackage{threeparttable}
\usepackage{cancel}
\usepackage{tikz}
\usetikzlibrary{decorations.pathmorphing}

\usetikzlibrary{decorations.markings}
\ifpdf
\usepackage[colorlinks=true,
linkcolor=black,
citecolor=black,
urlcolor=blue,
filecolor=blue,
pdfstartview=FitV,
pdftitle={},
pdfauthor={},
pdfsubject={Hydrodynamics},
pdfkeywords={hydrodynamics, AdS/CFT},
pdfpagemode=None,
bookmarksopen=true
]{hyperref}
\else
\usepackage[hypertex]{hyperref}
\fi
\usepackage{rotating}
\usepackage{rotating}
\makeatletter
\@addtoreset{equation}{section}

\makeatletter
\renewcommand\section{\@startsection {section}{1}{\z@}%
                                   {-3.5ex \@plus -1ex \@minus -.2ex}
                                   {2.3ex \@plus.2ex}%
                                   {\normalfont\large\bfseries}}
\renewcommand\subsection{\@startsection{subsection}{2}{\z@}%
                                     {-3.25ex\@plus -1ex \@minus -.2ex}%
                                     {1.5ex \@plus .2ex}%
                                     {\normalfont\bfseries}}

\def\sec#1{\S \;\ref{#1}}
\def\fig#1{Fig.\,\ref{#1}}

\title{{Magneto-Transport in a Chiral Fluid from Kinetic Theory}}

\author{Navid Abbasi$^{a}$\footnote{abbasi@ipm.ir},   \ Farid Taghinavaz$^{a}$\footnote{ftaghinavaz@ipm.ir}, \ Omid Tavakol$^{b}$\footnote{omid.t721@gmail.com }\\
\small{\emph{$^{a}$  School of Particles and Accelerators, Institute for Research in Fundamental Sciences (IPM)}}, \\
\small{\emph{P.O. Box 19395-5531, Tehran, Iran}}\\
\small{\emph{$^{b}$Department of Physics, Sharif University of Technology,}} \\
\small{\emph{P.O. Box 11365-9161, Tehran, Iran}} \\ [1mm]
}

\begin{document}

\setlength{\baselineskip}{16pt}
\begin{titlepage}
\maketitle

\vspace{-36pt}

\begin{abstract}
	We argue that in order to study the magneto-transport in a relativistic Weyl fluid, it is needed to take into account the associated quantum corrections, namely the side-jump effect, at least to second order. To this end, we impose Lorentz invariance to a  system of free Weyl fermions in the presence of the magnetic field and find the second order correction to the energy dispersion. By developing a scheme to compute the integrals in the phase space, we show that the mentioned correction has non-trivial effects on the thermodynamics of the system. Specifically, we compute the expression of the negative magnetoresistivity in the system from the enthalpy density in equilibrium.
 Then in analogy with Weyl semimetal, in the framework of the chiral kinetic theory and under the relaxation time approximation,  we explicitly compute the magneto-conductivities, at low temperature limit ($T\ll \mu$).  We show that the conductivities obey a set of Ward identities which follow from the generating functional including the Chern-Simons part.

  \end{abstract}
\thispagestyle{empty}
\setcounter{page}{0}
\end{titlepage}

\renewcommand{\baselinestretch}{1}  
\tableofcontents
\renewcommand{\baselinestretch}{1.2}  
\section{Introduction}
Studying chiral effects in quantum many body systems has attracted much interest during recent years.  It has in fact opened a window to macroscopically observe the anomalies of the microscopic quantum field theory.
In the hydrodynamic limit, such macroscopic manifestation occurs through the anomalous transport \cite{Son:2009tf}.  Although the main idea comes from the gauge/gravity computations, specifically form the fluid/gravity correspondence \cite{Banerjee:2008th,Erdmenger:2008rm},  the anomalous transport has also been directly studied in the field theory \cite{Landsteiner:2012kd}.

Since the anomalous transport is basically non-dissipative \cite{Kharzeev:2011ds}, another way to explore it is to study the thermodynamic  equilibrium in the system. In  \cite{Banerjee:2012iz,Jensen:2012jh} the dependence of the anomalous transport coefficients on the triangle anomaly has been found via constructing the equilibrium partition function. As shown in \cite{Jensen:2012kj}, the mixed gauge-gravitational anomaly coefficient may contribute to the anomalous transport, too, even in flat space-time.  In fact, the mixed gauge-gravitational anomaly contributes to the coefficients at two orders of derivatives lower than what expected from the equations of motion. Such contribution had been observed in free field theory of fermions \cite{Landsteiner:2011cp,Loganayagam:2012pz} as well as in  gauge/gravity duality \cite{Landsteiner:2011iq} \footnote{The mixed gauge gravitational anomaly has been realized as the origin for some interesting transport phenomena both in high energy and condensed matter physics \cite{Yamamoto:2015ria,Abbasi:2015saa,Chernodub:2015gxa,Chernodub:2018era,Huang:2018aly}.}. 

In an ideal Weyl gas, the anomalous transport has been related to the continuous injection of chiral states and their subsequent adiabatic flow
driven by vorticity \cite{Loganayagam:2012pz}.  This injection of states is shown to be corresponded with the flux of the Berry curvature through the Fermi surface in a Fermi liquid \cite{Son:2012wh}.  Putting a Berry monopole in the origin of momentum space, then a non-equilibrium  kinetic equation can be derived for the classical massless Weyl particles \cite{Stephanov:2012ki}. Using this so-called chiral kinetic theory, the anomalous transport can be studied beyond the hydrodynamic limit, specifically, one shows that the Berry curvature leads directly to the chiral magnetic effect.
The effect of Berry curvature on the dissipative transport has been also studied. In \cite{Weylsemimeta:Son}, it is shown that how  the Berry flux through the Fermi surface in a Weyl metal
gives rise to a large negative magnetoresistance, quadratically depending on the magnetic field.

On the other hand, it has been shown that in the presence of magnetic field, the Lorentz invariance dictates the energy dispersion of Weyl fermions gets a spin-magnetic correction \cite{Son:2012zy,Chen:2014cla}. This correction is actually a quantum correction and is necessary for showing the Lorentz invariance of the action of the massless spin-$\frac{1}{2}$ particles. The latter is realized by a modification of the Lorentz transformations; imposing a shift orthogonal to the boost vector and the particle momentum (side-jump effect). This ensures the angular momentum conservation in particle collisions.

 In the computation of the magneto resistance in the non-relativistic Weyl fluid in \cite{Weylsemimeta:Son}, the quantum correction to the energy dispersion has not been considered. As the first part in the current paper, we will compute the magneto-conductivities (including the magneto-resistance) in a relativistic Weyl fluid. To this end, we argue that one must take into account the second order correction to the energy dispersion. The reason is that this effect itself quadratically depend on the magnetic field, so the second order correction which is itself quadratic in the magnetic field unavoidably contributes to it. However, due to the form of the correction, some infrared divergences appear in the computations. By proposing a scheme for regulating the divergences, we will analytically perform the computations  in  the limit $\mu \gg T$.  While to first order in corrections, the thermodynamics of the system is not influenced, the second order correction turns out to have non-trivial effects on it. Specifically, we will discuss how one can compute the magnetoresistance in the system just by knowing the enthalpy density in equilibrium.

Recently, the observation of a positive longitudinal magnetothermoelectric
conductance in the Weyl semimetal NbP has been realized as the sign for the presence of the mixed gauge-gravitational anomaly in the condensed matter \cite{Gooth:2017mbd}.
Following this observation and to explore the effect through the chiral kinetic theory, we couple our system to a background temperature gradient and compute the thermoelectric coefficient as well. Using the linear response method, we then read the conductivities in the system under the relaxation time approximation.
 We consider that the dissipation effectively occurs at time scale $\tau$.\footnote{The DC transport in this case is like that of a Weyl semimetal. In a Weyl semimetal, in the regime that intervalley scattering time $\tau_{inter}$ is much larger than the intravalley scattering time $\tau_{intra}$ ($\tau_{inter}\gg \tau_{intra}$), the dissipation of momentum, energy and charge all are characterized with $\tau_{inter}$. This is due to the fact that the electron mean free path $\tau_{mfp}$ is essentially of the order of  $\tau_{intra}$ and in the  intervals of the order $t\sim\tau_{inter}\gg \tau_{mfp}$ the system is locally thermamilzed and therefore the transport occurs just through the  anomaly effects \cite{Weylsemimeta:Son}.}
Analogous to what was found in Weyl semimetal, we observe the positive longitudinal conductivity in the relativistic Weyl fluid.
Then by computing the heat current we confirm the validity of the Onsager reciprocal relation. We also compute the thermal conductivity coefficient.

The interesting point with our results in the kinetic theory is that the value of each conductivity, e.g. the electric conductivity, turns out to be 6.25$\%$ less than its value in the absence of quantum correction of the energy dispersion. Since the energy correction is related to the side-jump, this simply shows that due to the side-jump effect, the time between the successive scatterings in the system may decrease on average.

On the other hand, 
the same  decrease in the value of all conductivities suggests that there might be some linear relations between them. In a $2+1$ dimensional (non-chiral) system, it has been shown that the latter actually happens. Using the Ward identities, the authors of  \cite{Hartnoll:2009sz,Herzog:2009xv} find a set of relations between the electrical conductivity $\sigma$, thermoelectric $\alpha$ and thermal conductivity $\kappa$ coefficients. 

In order to find the probable relations between conductivities in the anomalous system we do as the following. Considering the standard inflow mechanism \cite{Harvey:2005it,Jensen:2012kj}, we first specify a generating functional which generates the stress tensor and charge current, in the presence of the anomalies.
In a system which is covariant under gauge and diffeomorphism variations, the charge current and stress tensor are uniquely defined from the covariant generating functional of the system. In our case, however, due to presence of the anomalies, one can choose whether to work with  "consistent" or "covariant" currents \cite{Bardeen}.  
By coupling the system to an external weak electric field and a weak background thermal gradient, we identify the "covariant current" as the one which responds to the electric field \footnote{Let us denote that in contrast to this rigorous statement, most of the computations in the ccontext of the transport are performed by using the "consistent" current and stress tensor.} and specify the form of the heat current as well.
We then derive a set of  Ward identities between one- and two-point functions of the covariant current and covariant stress tensor at zero momentum limit $k\rightarrow 0$. 
Using them, we would find two constraint equations between the transport coefficients like those of \cite{Hartnoll:2009sz,Herzog:2009xv}.
As a check of our computations in kinetic theory, we will show that the associated conductivities obey the constraints obtained from the  Ward identities. 

This consistency check shows more clearly the importance of the second order corrections coming from the chiral kinetic theory. It is not hard to show that without considering them, the constraints between conductivities will no longer be satisfied.

Finally, as the last evidence in favor of our results in this paper we compare them with those obtained from the covariant hydrodynamic model of Weyl semimetals developed in \cite{Lucas:2016omy}. While the model developed in  \cite{Lucas:2016omy} is suitable for more general cases, in one special case, its authors have applied their results to a system of weakly interacting Weyl gas with weak intervalley scattering.
They have then found two constraint equations between magneto-conductivities. We show that our conductivities obey the constraints obtained in the mentioned paper. This consistency suggests another approach to derive the magneto electrical resistivity in our system. 
To this end, by comparing the constraints obtained from Ward identities with those obtained in  \cite{Lucas:2016omy}, we will be able to find the electrical resistivity from a first order differential equation, once the charge density in equilibrium is given.

In the rest of the paper we do as it follows. In next section (\sec{kinetic}),  we first derive the second order correction imposed by the side-jump effect in the kinetic theory. Using that, we then compute the stress tensor components as well as the charge density. We compute the magneto-conductivities in the $\mu\gg T$ limit. We end the section by comparing the results with those of a Weyl semimetal. In \sec{Ward} we first introduce the generating functional which generates the stress tensor and charge current in the presence of both chiral and mixed guage-gravitational anomalies. Then by deriving the associated Ward identities in the limit $k \rightarrow 0$, we find two constraints between the magneto-conductivities coefficients.  We end in \sec{conclusion} with concluding and giving some future directions.

\section{Transport in Chiral Kinetic Theory}
\label{kinetic}
In what follows we consider an ensemble of right-handed chiral fermions. Due to charge conjugation, we have to consider the anti particles as well. The latter are the left-handed chiral fermions with opposite charge. The kinetic equation for the above two species of particles is given by
\begin{equation}\label{kinetic_eq}
\frac{\partial n^{(e)}_\textbf{p}}{\partial t}+\dot{\textbf{x}}\cdot\frac{\partial n^{(e)}_{\textbf{p}}}{\partial \textbf{x}}+\dot{\textbf{p}}\cdot\frac{\partial n^{(e)}_\textbf{p}}{\partial\textbf{p}}=I_{coll}\{n^{(e)}_{\textbf{p}}\},
\end{equation}
where from chiral kinetic theory we may write  \cite{Stephanov:2012ki}
\begin{eqnarray}\label{equation-Berry}
\sqrt{G}\dot{\textbf{x}}&=&\frac{\partial \epsilon_{\textbf{p}}}{\partial \textbf{p}}+e\, \textbf{E}\times \boldsymbol{\Omega}_{\textbf{p}} + e\, \textbf{B} \left(\frac{\partial \epsilon_{\textbf{p}}}{\partial \textbf{p}}\cdot\,  \boldsymbol{\Omega}_{\textbf{p}} \right),\\
\sqrt{G}\dot{\textbf{p}}&=& e\, \textbf{E}+ e \, \frac{\partial \epsilon_{\textbf{p}}}{\partial \textbf{p}} \times \textbf{B} + e^{2}\,\,  \boldsymbol{\Omega}_{\textbf{p}} \left(\textbf{E}\cdot\, \textbf{B}\right),
\end{eqnarray}
with $G=(1+e \textbf{B}\cdot \boldsymbol{\Omega}_{\textbf{p}})^2$ and 
\begin{equation}\label{epsilon_son}
\epsilon(\textbf{p})=\text{p}- \lambda\,e\,\frac{\textbf{B}\cdot\textbf{p}}{\text{p}^2},\,\,\,\,\,\,\,\,\boldsymbol{\Omega}_{\textbf{p}}=sgn( e)\lambda\frac{\boldsymbol{\hat{\text{p}}}}{\text{p}^2}.
\end{equation} 
In the expressions given above $\lambda=\pm1/2$ is the helicity associated with the right- and left-handed particles.
In the following we will be interested in the case in which the system is coupled to an external magnetic field in equilibrium. In the subsequent subsections, we would like to study the magneto-transport in the framework of chiral kinetic theory
\subsection{Relativistic corrections to energy dispersion of Weyl particles}
\label{central_result}
Relativistic invariance in a physical system forces the energy-momentum tensor of the system to be symmetric in every arbitrary Lorentz frame.  Let us consider the rest frame of the system which is in our present case is the laboratory frame as well. The above statement then says that the energy flux density $T^{i0}$ must be equal to the momentum density $T^{0i}$ in this frame. In what follows we first derive the corresponding expression  for these two objects.
Let us rewrite the energy density as 
\begin{equation}\label{energy_density}
T^{00}\equiv\epsilon=\int\frac{d^3\text{p}}{(2\pi)^3}\sqrt{G}\,\,\epsilon(\textbf{p})\,n_{\mathbf{p}}.
\end{equation}
Then we multiply equation \eqref{kinetic_eq} by $\sqrt{G} \epsilon(\textbf{p})$
and afterwards, integrate over $\textbf{p}$. Since $\sqrt{G} \epsilon(\textbf{p})$ is a collision-invariant object, the integral of the right-hand-side of the kinetic equation vanishes, when summing over particles and anti-particle contributions \cite{Landau_10}. Considering \eqref{energy_density}, the integrated equation then takes the form $\partial _{t}T^{0 0}+\partial_{i}T^{i0}=\text{E}^i j^i$ with the following expression for the energy flux density
\begin{equation}\label{T0i}
T^{i0}=-\int \frac{d^3 \text{p}}{(2\pi)^3}\left[(\delta^{ij}+e\,\text{B}^i\Omega^j)\frac{\epsilon_{\textbf{p}}^2}{2}\frac{\partial n_{\textbf{p}}}{\partial \text{p}^j}+\epsilon^{ijk}\frac{\epsilon_{\textbf{p}}^2}{2}\Omega^j\frac{\partial n_{\textbf{p}}}{\partial \text{x}^k}\right].
\end{equation}
On the other hand, analogous to what is defined in the classical kinetic theory, the momentum density can be simply defined as 
\begin{equation}\label{Ti0}
T^{0i}\equiv \pi^i=\int\frac{d^3\text{p}}{(2\pi)^3}\sqrt{G}\,\,\tilde{\text{p}}_i\,n_{\mathbf{p}}.
\end{equation}
Here $\tilde{\text{p}}_i=\text{p}_i-sgn(e)\,\frac{\lambda}{2}\,\epsilon_{ijk}\text{p}_j \partial_{k}$ is the modified momentum in phase space \cite{Abbasi:2017tea}. In our case however, the system is non-rotational and so in both the formula of $\tilde{\text{p}}$ and \eqref{T0i} the spacial partial derivative, namely $\partial_k$, vanishes in the equilibrium. Now by equating \eqref{T0i} with \eqref{Ti0} and integrating by part in \eqref{T0i}, we arrive at  
\begin{equation}\label{Lorentz_invariance}
T^{0i}=T^{i0},\,\,\,\,\,\rightarrow\,\,\,\,\,\,(\delta ^{ij}+e \text{B}^i \Omega_{\boldsymbol{p}}^j)\epsilon(\textbf{p})\frac{\partial \epsilon(\textbf{p})}{\partial \text{p}^j}=(1 +e \textbf{B}\cdot \boldsymbol{\Omega}_{\textbf{p}})\text{p}^i.
\end{equation}
As shown in 		\cite{Son:2012zy}, the above Lorentz invariance condition implies that the energy dispersion of particles in phase space gets correction due to spin-magnetic coupling. In the mentioned paper, the corresponding correction has been found to first order in the magnetic field.
According to our discussion in the introduction, in order to compute the magneto-conductivities, we have to find the second order correction to the energy dispersion as well. To this end, 
we take the following ansatz 
\begin{equation}\label{energy_correction}
\epsilon(\textbf{p})=\text{p}+\gamma_1(\text{p}) \,\textbf{B}\cdot \textbf{p}+ \gamma_2(\text{p})\,(\textbf{B}\cdot \textbf{p})^2.
\end{equation}
To find the two unknown functions $\gamma_1(\text{p})$ and $\gamma_2(\text{p})$, we insert the above ansatz in the equation \eqref{Lorentz_invariance} which leads to the two following equation
\begin{equation}\label{equation}
F(\text{p})\,(\textbf{B}\cdot \textbf{p})^2\hat{\text{p}}_i+\,G(\text{p})\,(\textbf{B}\cdot \textbf{p})\,\text{B}_i=\,0
\end{equation}
with $F$ and $G$ being as the following
\begin{eqnarray}
F(\text{p})&=&\text{p} \gamma_2'(\text{p})+\gamma_2(\text{p})+\gamma_1(\text{p}) \gamma_1'(\text{p})\\
G(\text{p})&=&\,2 \gamma_2(\text{p})+\frac{1}{ \text{p}^2}\gamma_1(\text{p})+\frac{1}{2 \text{p}}\gamma_1'(\text{p})+\gamma_1(\text{p})^2.
\end{eqnarray}
Since $\textbf{B}$ is independent of $\textbf{p}$, for \eqref{equation} to be held, it is needed both $F$ and $G$ functions vanish identically. Solving the coupled differential equations, we obtain \footnote{We work in the relativistic system of units with $\hbar=c=1$.}
\begin{equation}\label{central}
\gamma_1(\text{p})=-\frac{e}{2\text{p}^2},\,\,\,\,\,\,\,\,\,\boxed{\gamma_2(\text{p})=-\frac{e^2}{8\text{p}^5}}.
\end{equation}
While $\gamma_1(\text{p})$ was already found in \cite{Son:2012zy}, the $\gamma_2(\text{p})$ is our first new result in the current paper.
We will make clear the importance of such corrections in the next subsections, when computing the conductivities in a Weyl fluid.
\subsection{The effect of the quantum corrections on thermodynamics}
\label{thermo_quantities}
In this subsection, using the modified energy dispersion of Weyl particles, we compute some thermodynamic quantities in a thermal system of such particles, in the presence of a weak background magnetic field.
The equilibrium  distribution function for fermionic particles and anti-particles is simply given by
\begin{equation}
\tilde{n}_{\textbf{p}}^{(e)}=\frac{1}{e^{\beta \left(\epsilon(\text{p})-\frac{}{} \hat{e}\mu\right)}+1}.
\end{equation}
where  $\hat{e}$ denotes $sgn(e)=\pm1$ corresponding to particles and anti-particles, respectively. As we found in previous subsection, the energy dispersion, $\epsilon(\textbf{p})$, to second order in the quantum corrections is given by
\begin{equation}\label{energy_correction_second}
\epsilon(\textbf{p})=\text{p}-e\frac{\textbf{B}\cdot \textbf{p}}{2\text{p}^2}-e^2\frac{(\textbf{B}\cdot \textbf{p})^2}{8\text{p}^5}
\end{equation}
To perturbativley perform the computations, we expanded the equilibrium distribution function  to the same order
\begin{equation}\label{n_eq_expanded}
\tilde{n}_{\textbf{p}}^{(e)}=\tilde{n}_{\textbf{p}}^{(e)}\big|_{\epsilon(\textbf{p})=\text{p}}-\left(e\frac{\textbf{B}\cdot \textbf{p}}{2\text{p}^2}+e^2\frac{(\textbf{B}\cdot \textbf{p})^2}{8\text{p}^5}\right)\frac{\partial\tilde{n}_{\textbf{p}}^{(e)}}{\partial \epsilon}\big|_{\epsilon(\textbf{p})=\text{p}} +e^2\frac{(\textbf{B}\cdot \textbf{p})^2}{4\text{p}^4}\,\,\frac{\partial^2\tilde{n}_{\textbf{p}}^{(e)}}{\partial \epsilon^2}\big|_{\epsilon(\textbf{p})=\text{p}}.
\end{equation}

We use the above distribution function to compute the thermodynamic quantities. 
Let us start by computing the anomalous currents equilibrium.
It is well-known that in a chiral fluid, there are energy and charge currents in the equilibrium. These current are purely anomalous and in a fermionic system, they will no longer flow if the Dirac equation has no zero modes \cite{Zakharov:2012vv}. Taking \eqref{n_eq_expanded}, we can simply compute the energy and charge currents in the direction of magnetic field in a fermionic system with massless fermions. The coefficients, as one expects, are the anomalous transport coefficients in the Laboratory frame \cite{Landsteiner:2016led}, namely
\begin{align}\label{electric_current}
\boldsymbol{J}&=\sum_{e}\int_{\textbf{p}} e \sqrt{G}\,\dot{\textbf{x}}\,\,\tilde{n}_{\textbf{p}}^{(e)}\,\,\,\,\,\,\,\,\rightarrow\,\,\,\,\,\,\,\,\,\,\sigma_{\text{B}}=\frac{J_{\parallel}}{e\text{B}}=\,\frac{e\mu}{4 \pi^2}, \\
T^{0i}&=\sum_{e}\int_{\textbf{p}}\sqrt{G}\,\,\text{p}_i\,n_{\mathbf{p}}^{(e)}\,\,\,\,\,\,\,\,\,\,\rightarrow\,\,\,\,\,\,\,\,\,\,\sigma_{\text{B}}^{\epsilon}=\frac{T^{0 \parallel}}{e\text{B}}=\,\left(\frac{1}{3}+\frac{2}{3}\right)\left(\frac{\mu^2}{8 \pi^2}+\frac{T^2}{24}\right).
\end{align}
Here sum is over particle and anti-particle contributions. The integral $\int_{\textbf{p}}=\frac{d^3\textbf{p}}{(2\pi)^3}$ is performed over the allowed regions in the momentum space (see below).
The splitting of the fractional factors in front of $\sigma_{\text{B}}^{\epsilon}$ has been made for the following clarification. If one took the energy dispersion simply as $\epsilon(\textbf{p})=\text{p}$, he would obtain just $1/3$ of the total energy current. The additional $2/3$ contribution comes from the first order correction in \eqref{energy_correction_second}. The same situation was found for the coefficient of the  chiral vortical effect in a rotating system of chiral fermions \cite{Chen:2014cla}.  Let us denote that up the third order in the magnetic field, the second order correction in \eqref{energy_correction_second} does not contribute either to $\sigma_{\text{B}}$  or to $\sigma_{\text{B}}^{\epsilon}$.

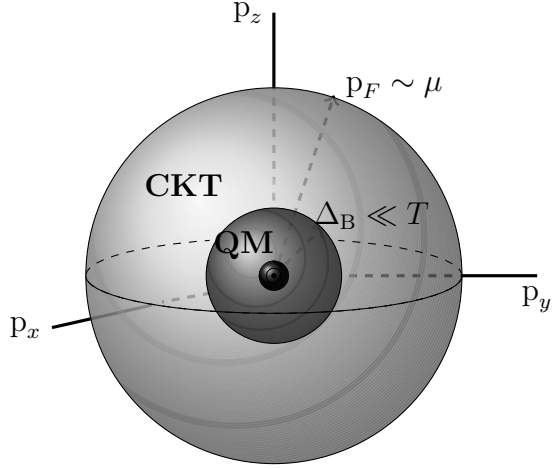
\begin{figure}
	\centering
	\begin{tikzpicture}
	\draw[very thick] (2.5,0) -- (3.5,0) coordinate (yaxis);
	\draw[very thick,dashed,->] (0,0) -- (.65,.65) coordinate ;
	\draw[very thick,dashed,->] (0,0) -- (0.8,2.4) coordinate ;
	\draw[very thick,dashed] (1,0) -- (2.5,0) coordinate ;
	\draw[very thick,dashed] (0,1) -- (0,2.5) coordinate ;
	\draw[very thick] (0,2.5) -- (0,3.5) coordinate (zaxis);
	\draw[very thick,dashed] (-.84,-.23) -- (-1.49,-0.35) coordinate ;
	\draw[very thick] (-1.62,-0.39) -- (-2.95,-0.69) coordinate (xaxis);
	\shade[ball color = gray!40, opacity = 0.4] (0,0) circle (2.5cm);
	\draw (0,0) circle (2.5cm);
	\shade[ball color = black!70, opacity = 0.4] (0,0) circle (.9cm);
	\draw (0,0) circle (0.9cm);
	\shade[ball color = black!120, opacity = 0.4] (0,0) circle (.2cm);
	\draw (0,0) circle (0.1cm);
	\draw[dashed] (2.5,0) arc (0:70:2.5 and 0.5);
	\draw[dashed] (2.5,0) arc (360:110:2.5 and 0.5);
	\draw (2.5,0) arc (360:180:2.5 and 0.5);
	\fill[fill=black] (0,0) circle (1pt);
	\node[below] at (yaxis) {$\text{p}_y$};
	\node[left] at (zaxis) {$\text{p}_z$};
	\node[left] at (xaxis) {$\text{p}_x$};
	\node at (1.3,.8) {$\Delta_{\text{B}}\ll T$};
	\node at (-1.2,1.2) {\textbf{CKT}};
	\node at (-0.4,0.4) {\textbf{QM}};
	\node at (1.6,2.5) {$\text{p}_{F}\sim \mu$};
	\end{tikzpicture}
	\caption{In the low temperature limit $T \ll \mu$, the quasi-particles occupy the states inside the Fermi sphere. When considering the interaction with Berry monopole located at the origin, the states within the inner sphere with radius $|\textbf{p}|=\Delta_{B}$ are excluded. In this region, the quantum mechanical effects are dominant.  The outer sphere shows the Fermi surface in the limit $\mu\gg T$. While the kinetic theory works for $ |\textbf{p}| \lesssim \mu$, chiral kinetic theory is a valid picture in the range $\Delta_{B}  \lesssim |\textbf{p}| \lesssim \mu$.}
	\label{fig_sphere}
\end{figure}

One important place wherein the second order correction of the energy dispersion (given in \eqref{energy_correction_second}) makes a non-trivial role is the diagonal components of the energy momentum tensor $T^{\mu \nu}$. Let us follow the issue by computing the $T^{00}$ component.
The energy density in equilibrium is given by
\begin{align}\label{T00}
T^{00}\equiv&\,\epsilon=\sum_{e}\int_{\textbf{p}}\sqrt{G}\,\,\epsilon(\textbf{p})\,\tilde{n}_{\mathbf{p}}^{(e)}\\\nonumber
=&\int_{0}^{+\infty}\frac{d\text{p}}{2 \pi^2}\frac{1}{1+e^{\beta(\text{p}-\mu)}}\left[\text{p}^3-\frac{e^2\text{B}^2}{8 \text{p}}+\frac{e^2\text{B}^2}{24 T^2}\frac{e^{\beta(\text{p}-\mu)}(T+\text{p})+e^{2\beta(\text{p}-\mu)}(T-\text{p})}{(1+e^{\beta(\text{p}-\mu)})^2}\right]
\end{align}
Due to presence of the $1/\text{p}^5$ term in	\eqref{n_eq_expanded}, the integral in \eqref{T00} is obviously IR divergent (see the second term in the brackets.). In fact, what may remove the divergence is that the magnitude of the momentum in the phase space has to be bounded from below. Let us recall that in order to put the chiral fermions in the framework of the kinetic theory, we have already assumed that the value of the background magnetic field to be such small that the particles move on classical trajectories \cite{Stephanov:2012ki}. For the latter to be acquired, the  necessary condition is
$\sqrt{e\text{B}}\ll \text{p}$ which simply states that the momentum integrals have to be regularized by considering an IR cut-off $\Delta_{\text{B}} \lesssim \text{p}$.
Let us recall that according to expansion given by \eqref{n_eq_expanded}, we perform the computations perturbatively, in powers of $\text{B}$. Our scheme is to treat with the non-magnetic parts of the integrals differently compared to the magnetic parts which come from the interaction of the Berry flux with fermions.
For the former in $\mu \gg T$ limit, we consider the quasi-particles with all momenta inside the Fermi sphere while for the latter, we restrict the computations to be performed for the quasi-particles with momenta higher than the cut-off inside the sphere (See \fig{fig_sphere}.). The scheme is basically originated from the derivation of the chiral kinetic theory in \cite{Stephanov:2012ki}. It turns out that this scheme leads to physical results.

Since the cut-off arises due to the magnetic field,  in $T\ll\mu$ limit then it is reasonable to take the cut-off as being of the order $\sqrt{eB}\ll \Delta_{\text{B}} \ll T$.  Therefore, the order of scales may be written as
\begin{equation}\label{regime_setup}
\sqrt{e\text{B}}\ll \Delta_{\text{B}}\ll T \ll \mu.
\end{equation}
This is also in complete correspondence with the hydrodynamic limit, which we arrive at later on. Under the above considerations we find 
\begin{align}\label{T_00}
T^{00}\equiv\epsilon&=T^4 \left(\frac{\mu^4}{8 \pi ^2 T^4}+\frac{\mu^2}{4 T^2}+\frac{7 \pi ^2}{120}\right)+
\frac{e^2\text{B}^2}{24 \pi^2}-\left(log \frac{\mu}{\Delta_{\text{B}}}-\frac{\pi^2}{6}\frac{T^2}{\mu^2}+\,O(\frac{T^4}{\mu^4})\right)\frac{e^2\text{B}^2}{16 \pi^2}
\end{align}
where the term including $\Delta_{\text{B}}$  appears to cancel out the contribution of the excluded region in \fig{fig_sphere} (See Appendix \ref{regular} for details.). 
 However, the $\log$ term
is comparable with the leading correction. We will see in the following that this dependence on the cut-off will vanish in the enthalpy density and so do happen in all the conductivities.

Before proceeding to compute the other diagonal components of the stress tensor, let us first consider the thermodynamic pressure. From the equilibrium partition function and by using the method developed in Appendix \ref{regular} we find
\begin{align}\label{pressure}
p&=T \sum_{e}\int_{\textbf{p}}\sqrt{G}\,\,log\left(1+\frac{}{}e^{-\beta \left(\epsilon(\text{p})-\frac{}{} \mu\right)}\right)\\ \nonumber
&=\int_{0}^{+\infty}\frac{d \text{p}}{2\pi^2}\text{p}^2T \log (1+e^{-\beta(\text{p}-\mu)})+\int_{\Delta}^{+\infty}\frac{d \text{p}}{2\pi^2}\left[\frac{e^2\text{B}^2}{8 \text{p}}\frac{1}{1+e^{\beta(\text{p}-\mu)}}+\frac{e^2\text{B}^2}{24 T}\frac{e^{\beta(\text{p}-\mu)}}{(1+e^{\beta(\text{p}-\mu)})^2}\right]\\\nonumber
&=T^4 \left(\frac{\mu^4}{24 \pi ^2 T^4}+\frac{\mu^2}{12 T^2}+\frac{7 \pi ^2}{360}\right)
+\frac{e^2\text{B}^2}{48 \pi^2}+\left(log \frac{\mu}{\Delta_{\text{B}}}-\frac{\pi^2}{6}\frac{T^2}{\mu^2}+\,O(\frac{T^2}{\mu^2})\right)\frac{e^2\text{B}^2}{16 \pi^2}.
\end{align}

Interestingly, while both energy density and pressure get logarithmic correction in the presence of the magnetic field, the sum of them, namely the enthalpy density, reads
\begin{equation}\label{enthalpy_CKT}
w=\,\epsilon+p=\,T^4 \left(\frac{\mu^4}{6 \pi ^2 T^4}+\frac{\mu^2}{3 T^2}+\frac{7 \pi ^2}{90}\right)+\,\frac{e^2\text{B}^2}{16 \pi^2}\left(1+O(\frac{T^5}{\mu^5})\right)
\end{equation}
Note that in all expressions given above (and also those that come below in the current subsection), we give each quantity, like the energy density, pressure and ... , as the sum of two  parts; the first part is the exact form of the quantity in the absence of the magnetic field; the second part reads the quadratic quantum correction of the magnetic field to the quantity in the limit $T\ll \mu$.
This is the regime in which the computations are analytically performed. Interestingly, \eqref{regime_setup} insures that all terms in the first part of each quantity are leading compared to the second part terms. So, we keep all the first part contributions without truncating them in the $\frac{T}{\mu}$ expansion. 

Let us now turn back to the computation of the spacial diagonal components of the stress tensor. 
In  a magnetic system, one naturally expects to see difference between the value of the diagonal components of the stress tensor in the direction of the magnetic field compared to those of the transverse directions.
To make the difference clear, we now compute the diagonal components of the stress tensor. 
In a non-rotating equilibrium state, like the one under study in the current paper, we may write \cite{Son:2012zy}
\begin{equation}\label{}\nonumber
T^{ij}=-\sum_{e}\int_{\textbf{p}}\text{p}_i\,\left(\epsilon(\textbf{p})(\delta^{jk}+e\text{B}^j \Omega^k)\,\frac{\partial\tilde{n}_{\mathbf{p}}^{(e)}}{\partial \text{p}^k}+ e\epsilon ^{jkl}\, \Omega^k \text{E}^l\, \tilde{n}_{\mathbf{p}}^{(e)}\right)-  \delta^{ij} \epsilon.
\end{equation}
Considering the magnetic field being in the 3-direction, $T^{33}$  turns out to be exactly equal to the thermodynamic pressure obtained from the partition function \eqref{pressure} and  $T^{22}=T^{11}$ reads
\begin{align}\label{T_11}
T^{22}=T^{11}=&\,\,\frac{1}{3T}\int_{0}^{+\infty}\frac{d \text{p}}{2 \pi^2}\frac{\text{p}^4\,e^{\beta(\text{p}-\mu)}}{(1+e^{\beta (\text{p}-\mu)})^2}\\\nonumber
&+\frac{e^2\text{B}^2}{120 T^3}\int_{\Delta}^{+\infty}\frac{d \text{p}}{2 \pi^2}\frac{\text{p}\,e^{\beta(\text{p}-\mu)}}{(1+e^{\beta (\text{p}-\mu)})^3}\left[\text{p}\frac{(e^{\beta(\text{p}-\mu)}-1)^2-2e^{\beta(\text{p}-\mu)}}{1+e^{\beta (\text{p}-\mu)}}+3T(e^{\beta (\text{p}-\mu)}-1)\right].
\end{align}
Performing the above integrals one finds 
\begin{equation}\label{T_11_22_33}
T^{11}=T^{22}=T^{33}-\frac{e^2\text{B}^{2}}{24\pi^{2}}=p-\frac{e^2\text{B}^{2}}{24\pi^{2}}.
\end{equation}

Another thermodynamic quantity which is influenced by the second order correction of energy dispersion is the charge density
\begin{align}\label{density}\nonumber
n=\sum_{e}\int_{\textbf{p}}  e\, \tilde{n}_{\textbf{p}}^{(e)}\,\,
&=\int_{0}^{\infty}\frac{1}{2\pi^2}\frac{\text{p}^2}{1+e^{\beta(\text{p}-\mu)}}+\,\frac{e^2\text{B}^2}{24 T^2}\int_{\Delta}^{\infty}\frac{1}{2\pi^2}\left[\frac{3T}{ \text{p}}\frac{e^{\beta(\text{p}-\mu)}}{(1+e^{\beta(\text{p}-\mu)})^2}+\frac{e^{\beta(\text{p}-\mu)}(e^{\beta(\text{p}-\mu)}-1)}{(1+e^{\beta(\text{p}-\mu)})^3}\right]\\
&=\,T^3 \left(\frac{\mu^3}{6 \pi ^2 T^3}+\frac{\mu }{6 T}\right)+\frac{e^2\text{B}^2}{16 \pi^2 \mu}\left(1+\frac{\pi^2T^2}{3\mu^2}+\,O(\frac{T^4}{\mu^4})\right).
\end{align}
As before, the last term can be drooped.
One can also simply check that this relation might be obtained via $n=\big(\partial p/\partial \mu\big)_T$ by using \eqref{pressure}.
The appearance of the $\text{B}^2$ contributions in the thermodynamic quantities associated with Weyl fermions, although observed as a new result in the current paper, is not surprising. The quasi-particles in the system are weakly interacting with the magnetic field and consequently the system becomes magnetized. The situation is similar to what is studied in the magnetohydrodynamics \cite{Hernandez:2017mch}. We come back to this point in the next subsection.

Before ending this subsection, let us make a point about the entropy density in equilibrium. Considering the thermodynamic relation $\epsilon+p =Ts +\mu n$ and by using the thermodynamic quantities found above, we arrive at
\begin{equation}
s=T^3\left(\frac{7\pi^2}{90}+\frac{\mu^2}{6T^2}\right)-\frac{e^2\text{B}^2 T}{24\pi^2 \mu^2}\left(1+O(\frac{T^2}{\mu^2})\right)
\end{equation}
which simply shows that 
the presence of magnetic field has decreased the entropy density in the system. The same situation was  observed in a strongly coupled system before. In \cite{DHoker:2009ixq}, it has been shown that in a $\mathcal{N}=4$ SYM gauge theory, the presence of magnetic field reduces the entropy density. It suggests that the decrease in the entropy density due to the magnetic effects might be a universal behavior in chiral systems.
\subsection{More about the thermodynamics of the system: a physical prediction}
In the previous subsection in the framework of kinetic theory, we computed the thermodynamic quantities of the system of free massless fermions with considering the second order quantum corrections. In order for  the chiral kinetic theory be applicable, we demanded the magnetic field be sufficiently weak, $\text{eB}\ll T^2$.
In the language of magneto-thermodynamics developed in \cite{Hernandez:2017mch} our system is described by the following free energy density 
\begin{equation}
\mathcal{F}=\,p(T,\mu,\textbf{B}^2)+\,O(\partial^3).
\end{equation}
While in \cite{Hernandez:2017mch} with the assumption $B\sim O(1)$ the free energy $p(T,\mu, B^2)$ would be a zero derivative object, in our construction it is in fact as a corrected quantity to second order in derivatives. Interestingly as we showed, this is exactly the order to which we have to keep terms to study the magneto-transport. The stress tensor and charge current in equilibrium are given by 
\begin{equation}\label{T_MHD_equ}
T^{\mu \nu}=(\epsilon+\,\Pi)u^{\mu}u^{\nu}+\Pi\, \eta^{\mu \nu}+\alpha_{\text{B}\text{B}}\left(B^{\mu}B^{\nu}-\frac{1}{3} \Delta^{\mu \nu}B^{2}\right),\,\,\,\,\,\,\,\,J^{\mu}= n\, u^{\mu}
\end{equation}
with $\Pi=p-\frac{2}{3}\alpha_{\text{BB}}B^2$.
Here $\epsilon$, $p$, $\alpha_{\text{B}\text{B}}$ and $n$ are functions of $(T, \mu, B^2)$ in general.  $\alpha_{\text{B}\text{B}}(T, \mu, B^2)$ is the magnetic susceptibility. In the above expressions, $u^{\mu}$  is the velocity of the rest frame  of the equilibrium state and $B^{\mu}$ is the magnetic field in the rest frame.
Using the definition $\Delta^{\mu\nu}=\eta^{\mu \nu}+u^{\mu}u^{\nu}$, equation \eqref{T_MHD_equ} is rewritten as
\begin{equation}\label{T_MHD_eq_alphaBB}
T^{\mu \nu}=\big(\epsilon+p-\alpha_{\text{BB}}B^2\big)u^{\mu}u^{\nu}+\big(p-\alpha_{\text{BB}}B^2\big)\eta^{\mu \nu}+\alpha_{\text{BB}}B^{\mu}B^{\nu}. 
\end{equation}
Then by taking $u^{\mu}=(1,0,0,0)$ and $B^{\mu}=(0,0,0,\text{B})$, one obtains
$T^{00}=\epsilon$ and $T{^{33}=p}$.
The energy and pressure satisfy the following relations
\begin{equation}\label{Gibbs_Duhem}
\epsilon+p= T (\partial p/\partial T)_{\mu}+\mu (\partial p/\partial \mu)_{T}
\end{equation}
and $n=(\partial p/\partial \mu)_{T}.$
Now as an example we consider the thermodynamics of the free fermionic system studied in the previous subsection. The mentioned system could be regarded as the magneto-thermodynamic state given above, however,  with a special equation of state and consequently with a specific $\alpha_{\text{B}\text{B}}$. In the following we discuss on two points in this system. First, we physically motivate that the $\text{B}^2$ dependence of the enthalpy density is in relation with the longitudinal magneto-conductivity. Then by constructing a covariant formula for $\alpha_{\text{BB}}$, we find its value for our system.

Taking $\epsilon$ and $p$ as given by \eqref{T00} and \eqref{pressure}, one can simply check that the relation \eqref{Gibbs_Duhem} identically holds. This can be regarded as a check for our thermodynamic computations. There is another point, however, with the $\text{B}^2$-dependent in \eqref{enthalpy_CKT}. This term has a nice relation with the electrical conductivity. We follow the interesting discussion on "chiral battery"
in \cite{Fukushima:2008xe} and explain the relation in the following.

Let us consider the system in the magnetic field \text{B} and at a chiral chemical potential $\mu$. As shown in \eqref{electric_current}, the magnetic field induces an electric current in the chiral system $J^{\parallel}=e^2\mu\text{B}/4 \pi^2$. This is in fact the statement of chiral magnetic effect. On the other hand if the conductivity of system is finite\footnote{A finite conductivity is always the sign for the presence of a microscopic scattering mechanism in the system. In the case of WSM, the latter may be related to inter-valley scattering in the momentum space\cite{Weylsemimeta:Son}.}, say $\sigma=\rho^{-1}$ with $\rho$ being the resistivity, according to the Ohm's law this current induces potential difference between the points, $V^{\parallel}= \rho\, J^{\parallel}$.
Since existence of potential difference is  equivalent to having an electric field $\text{E}$, one expects this field together with \text{B}, turn on the axial anomaly and decrease the density of chiral charges. Due to this anomalous non-conservation of chiral charges, the corresponding chemical potential, namely $\mu$, will no longer be constant. Let us take the time scale over which the chiral chemical potential approaches zero as $\tau$. If $\tau$ is much larger than all microscopic time scales\footnote{This is our basic assumption in the whole of this paper.}, then the rate of change of $\mu$ can be taken as constant being equal $\mu/\tau$ (up to corrections of order $1/\tau^2$). This is nothing but the electric field \text{E} discussed above, so we can write $e\text{E}=\mu/\tau$\footnote{Let us recall that the electric field induced due to the change in temperature and chemical potential is given by $e \textbf{E}=T \boldsymbol{\nabla}(\mu/T)$\cite{Lucas:2016omy}. When temperature is constants, it simplifies to $e \textbf{E}=\boldsymbol{\nabla}\mu$. }.  Considering the electrical conductivity as $\sigma_{L}$, the mean heat power produced in the system then would be $  \boldsymbol{J\cdot \textbf{E}}=(\sigma_{L} \text{E})\,\text{E}=\,\sigma_{L} \mu^2/e^2\tau^2$\footnote{The chiral magnetic effect current gradually decreases due to non-conversation of  chiral chemical potential $\mu$. Its mean value, namely $\boldsymbol{J}$, is the response to the constant electric field $\textbf{E}$ as $\boldsymbol{J}=\sigma_{L} \textbf{E}$. }. Thus the amount of heat produced in the time $\tau$, is 
$(\boldsymbol{J\cdot \textbf{E}})\tau=\, \sigma_{L}\, \mu^2/e^2\tau$.
When the pressure is constant, this heat is equivalent to the enthalpy density and gives rise to the $\text{B}^2$-dependent term in it. Considering \eqref{enthalpy_CKT}, one writes
\begin{equation}\label{prediction}
\sigma_{L} \frac{\mu^2}{e^2\tau}=\,\frac{e^2\text{B}^2}{16 \pi^2}\,\,\,\,\,\,\rightarrow\,\,\,\,\,\,\,\,\sigma_{L}=\,\frac{\tau e^4\text{B}^2}{16 \pi^2 \mu^2}.
\end{equation}
This is an interesting result about the magneto-conductivity, or inversely about the negative magnetoresistivity, which we obtained from thermodynamic arguments. 
In next subsections, we confirm this physical discussion via studying the linear response of the system to an external electric field.

Let us now compute the magnetic susceptibility $\alpha_{\text{BB}}$. 
We can take the longitudinal and transverse pressure, respectively as $p_{\parallel}=p$ and $p_{\perp}=p- \text{M}\, \text{B}$ where the magnetization vector is defined by $\textbf{M}= \alpha_{\text{BB}} \textbf{B}$
\cite{Huang:2011dc}. It is obvious that the magnetic susceptibility represents the relative difference between $p_{\parallel}$ and $p_{\perp}$. To find a covariant formula for $\alpha_{\text{BB}}$ one can find two operators acting on \eqref{T_MHD_eq_alphaBB} that  project out the transverse and longitudinal pressures 
\begin{align}
p&=\,p_{\parallel}=\,b_{\mu}b_{\nu}\,T^{\mu \nu}\\
p- \text{M}\, \text{B}&=\,p_{\perp}=\,\frac{1}{2}(b_{\mu}b_{\nu}-\Delta_{\mu\nu})T^{\mu \nu}
\end{align}
where $b^{\mu}=B^{\mu}/B$ \footnote{It should be noted that $B^{\mu}B_{\mu}=B^2$ and $B =|\textbf{B}|$.}. As a result we find
\begin{equation}
\alpha_{\text{BB}}\,\,B^2=\,\frac{1}{2}(3\,b_{\mu}b_{\nu}-\Delta_{\mu\nu})\,T^{\mu \nu}.
\end{equation}
Applying the above formula to the equilibrium state given below \eqref{T_MHD_eq_alphaBB} and by using  \eqref{pressure} and \eqref{T_11_22_33}, the susceptibility in the system of free massless fermions turns out to be as the following
\begin{equation}\label{alpha_BB}
\alpha_{\text{BB}}=\,\frac{e^2}{24 \pi^2}.
\end{equation}

In summary, in this subsection we explained how to describe the thermodynamics of the system of free massless fermions in the framework of magneto-thermodynamics developed in \cite{Hernandez:2017mch}. As an example to general arguments of the latter reference we showed that in our fermionic system, the $\text{B}^2$-dependent of the enthalpy density is related to the longitudinal electrical conductivity \eqref{prediction}. We also explicitly computed the magnetic susceptibility in the system \eqref{alpha_BB}.

\subsection{Dynamics towards equilibrium}
Let us suppose in a system of non-interacting Weyl fermions, the dissipating dynamics towards equilibrium is governed by a relaxation time approximation with parameter $\tau$
\begin{equation}\label{kinetic_eq_RTA}
\frac{\partial n^{(e)}_\textbf{p}}{\partial t}+\dot{\textbf{x}}\cdot\frac{\partial n^{(e)}_{\textbf{p}}}{\partial \textbf{x}}+\dot{\textbf{p}}\cdot\frac{\partial n^{(e)}_\textbf{p}}{\partial\textbf{p}}=-\frac{n^{(e)}_\textbf{p}-\tilde{n}^{(e)}_\textbf{p}}{\tau}.
\end{equation}
We are interested in a steady state case, i.e. $\omega=0$, wherein, the system is homogeneous as well. Under such considerations and in the presence of a weak magnetic field $B\ll T^2$\footnote{In this paper, we consider a low temperature system of chiral fermions, i.e. $T\ll \mu$.}, we use linear response theory to study the response of the system a probe electric field. So \eqref{kinetic_eq_RTA} can be written in the following linearized form
\begin{equation}\label{steady_state}
\frac{1}{\sqrt{G}}\left( e\, \textbf{E}+ e^{2}
\boldsymbol{\Omega}_{\textbf{p}} \left(\textbf{E}\cdot\textbf{B}\right)\right)\cdot\frac{\partial \tilde{n}^{(e)}_\textbf{p}}{\partial\textbf{p}}=-\frac{n^{(e)}_\textbf{p}-\tilde{n}^{(e)}_\textbf{p}}{\tau}\equiv -\frac{\delta n_{\textbf{p}}^{(\epsilon)}}{\tau}.
\end{equation}
Let us denote that $\tilde{n}^{(e)}_{\textbf{p}}$ is  the equilibrium distribution function, while  $n^{(e)}_{\textbf{p}}$ is the linear response of the system to the electric field fluctuation $\textbf{E}$. 

In the present case where the system is assumed to be uniform and time independent,    $\delta n^{(e)}_{\textbf{p}}= n^{(e)}_{\textbf{p}}-\tilde{n}^{(e)}_{\textbf{p}}$ in the RHS of \eqref{steady_state} can be written in terms of the hydrodynamic variables. In the simple case with just one single chirality in the system, the hydrodynamical variables are the three components of fluid velocity $u^{\mu}$, temperature $T$ and the chiral chemical potential $\mu$. One may write
\begin{equation}\label{Boltzaman_hydro}
\frac{\partial n^{(e)}_\textbf{p}}{\partial T}\delta T+ \frac{\partial n^{(e)}_\textbf{p}}{\partial \mu}\delta \mu+\frac{\partial n^{(e)}_\textbf{p}}{\partial \boldsymbol{u}}\cdot\delta \boldsymbol{u}=\,-\frac{\tau}{\sqrt{G}}\left( e\, \textbf{E}+ e^{2}
\boldsymbol{\Omega}_{\textbf{p}} \left(\textbf{E}\cdot\textbf{B}\right)\right)\cdot\frac{\partial \tilde{n}^{(e)}_\textbf{p}}{\partial\textbf{p}}.
\end{equation} 
By computing the moments of this equation then one finds the conservation equations for energy, momentum and charge in the hydrodynamic regime.

What we are going to do in the following is a little different from this point of view. We get $\delta n_{\textbf{p}}^{(\epsilon)}$ from \eqref{steady_state}  without entering any hydrodynamic variable. Using this, we compute the thermal and electrical conductivities in the system. 

\subsection{Transport from chiral kinetic theory}
\label{sec_transport}

In the presence of a background magnetic field we couple the system to weak electric field together with a weak temperature gradient. 
Then we compute the electric current as well as   heat current. They take the following form:
\begin{eqnarray}\label{j_E}
\boldsymbol{J}_{e}&=&\sigma \, \textbf{E}+T \alpha_1 \left(-\frac{\boldsymbol{\nabla}\text{T}}{\text{T}}\right),\\\label{j_th}
\boldsymbol{J}_{th}&=&T\alpha_2  \,\textbf{E}+ T\kappa \left(-\frac{\boldsymbol{\nabla}\text{T}}{\text{T}}\right).
\end{eqnarray}
In the relation above, $\sigma$ is the electrical conductivity and $\kappa$ is the thermal conductivity coefficient. The other coefficient, namely $\alpha_1=\alpha_2$, is the thermoelectric effect coefficient which is related to induction of an electric (or thermal) current as the response to the presence of a temperature gradient (or electric field) in the system. In what follows for simplicity we get $\boldsymbol{\zeta}\equiv-\boldsymbol{\nabla}T/T$.

Since  the temperature is being assumed to have a gradient in the system, it has to be well-defined as well in the whole of the system. This means that its variation should be such long-wavelength that one can locally define the temperature at each point in the system. This is simply acquired in the hydrodynamic limit.
So in order to enter the background temperature gradient, we limit the following discussion to a special case in which the equilibrium configuration of the system is a zero order hydrodynamic profile. The out of equilibrium distribution function is then given by 
\begin{equation}\label{n_x}
n^{(e)}_{\textbf{p}}= \tilde{n}^{(e)}_{\textbf{p}}(\textbf{x})+\delta n^{(e)}_{\textbf{p}}=\,\frac{1}{e^{\beta(\textbf{x})\left(\epsilon(\textbf{p})-\hat{e} \mu\right)}+1}+\delta n^{(e)}_{\textbf{p}}.
\end{equation} 
Let us recall that we would like to study the response of the system
with respect to the two external sources; first the external electric field which appears in $\delta n^{(e)}_{\textbf{p}}$ in the equation above. Second, a source of  temperature gradient which comes with the gradient of $\tilde{n}^{(e)}_{\textbf{p}}(\textbf{x})$. Let us elaborate on the latter. One may expand  $\tilde{n}^{(e)}_{\textbf{p}}$ around equilibrium state whose temperature is constant $T=1/\beta$. We obtain
\begin{equation}\label{n_p_x}
\tilde{n}^{(e)}_{\textbf{p}}(\textbf{x})=\,\tilde{n}^{(e)}_{\textbf{p}}+\,\frac{\partial  \tilde{n}^{(e)}_{\textbf{p}}}{\partial T} \,\textbf{x}\cdot\boldsymbol{\nabla} T=\,\tilde{n}^{(e)}_{\textbf{p}}+\big(\epsilon(\textbf{p})-\hat{e} \mu\big)\frac{\partial  \tilde{n}^{(e)}_{\textbf{p}}}{\partial \text{p}} \,\,\textbf{x}\cdot\boldsymbol{\zeta}
\end{equation}
The explicit dependence on $\textbf{x}$ will vanish once one finds $\frac{\partial n^{(e)}_{\textbf{p}}}{\partial \textbf{x}}$ in \eqref{kinetic_eq_RTA}.
In order to use the linear response theory in the presence of the above-mentioned sources, it is required that
\begin{equation}
\nabla \beta\,\tilde{n}^{(e)}_{\textbf{p}}\sim \beta^2\, \text{E} \,\tilde{n}^{(e)}_{\textbf{p}}\sim \delta n^{(e)}_{\textbf{p}} \ll n^{(e)}_{\textbf{p}}.
\end{equation}
Now, substituting \eqref{n_x} (with \eqref{n_p_x}) into \eqref{kinetic_eq_RTA}, we get new contributions from the second term of \eqref{kinetic_eq_RTA} in the LHS, even in the steady and uniform case. One writes
\begin{equation}\label{delta_n_E_T}
\delta n^{(e)}_{\textbf{p}}= -\frac{\tau }{\sqrt{G}}\left[\left(e\textbf{E}\cdot \textbf{v}_{\textbf{p}} +e^2(\boldsymbol{\Omega_{p}}\cdot \textbf{v}_{\textbf{p}})\frac{}{} \textbf{E}\cdot \textbf{B}\right)+(\epsilon(\textbf{p})-\hat{e}  \mu)\left(\boldsymbol{\zeta}\cdot \textbf{v}_{\textbf{p}} +e(\boldsymbol{\Omega_{p}}\cdot \textbf{v}_{\textbf{p}})\frac{}{} \boldsymbol{\zeta}\cdot \textbf{B}\right)\right]\frac{\partial  \tilde{n}^{(e)}_{\textbf{p}}}{\partial \text{p}}
\end{equation}
with the group velocity of the quasi particles being as
$\textbf{v}_{\textbf{p}}=\frac{\partial \epsilon}{\partial \textbf{p}}$.
Having found the deviation from the equilibrium, in the two following parts in this section, we compute the electric and thermoelectric conductivities.

We first neglect the temperature gradient in the system and consider only the response of the system to the external electric source.
According to \eqref{steady_state}, the deviation from equilibrium has two parts; first the Ohm contribution $\delta n^{(e)}_{\text{O}}$, which is simply due to work done on the charged particles in the system by the electric field. The second,  $\delta n^{(e)}_{\text{A}}$ is due to the anomaly. One may write
\begin{equation}\label{delta_n_O_A}
\delta n^{(e)}_{\textbf{p}}=\,\delta n^{(e)}_{\text{O}}+\delta n^{(e)}_{\text{A}}
= \,-\frac{\tau e}{\sqrt{G}}\textbf{E}\cdot \textbf{v}_{\textbf{p}} \frac{\partial  \tilde{n}^{(e)}_{\text{p}}}{\partial p}-\frac{\tau e^2}{\sqrt{G}}(\boldsymbol{\Omega_{p}}\cdot \textbf{v}_{\textbf{p}})\frac{}{} \textbf{E}\cdot \textbf{B}\,\frac{\partial  \tilde{n}^{(e)}_{\text{p}}}{\partial p}
\end{equation}
Considering \eqref{electric_current}, we multiply \eqref{delta_n_O_A} with $e\sqrt{G}\boldsymbol{\dot{x}}$ and then integrate over the momentum space. To proceed, it is also needed to use  \eqref{equation-Berry}. For clarifying, in the following, we bring the detailed computations in this case. The electric current parallel to the magnetic field, linearized in $\text{E}$ is given by
\begin{align}\label{j_z_ohm_CME}\nonumber
J^{\parallel}_{e}=&-\sum_{e}\int_{\boldsymbol{p}}\tau\,e^2 \, \dot{\textbf{x}}_{\parallel}\,\left( \frac{}{} \textbf{E}\cdot \textbf{v}_{\textbf{p}}+\,e\,(\boldsymbol{\Omega_{p}}\cdot \textbf{v}_{\textbf{p}})\frac{}{} \textbf{E}\cdot \textbf{B}\right)\, \frac{\partial  \tilde{n}^{(e)}_{\textbf{p}}}{\partial \text{p}}
\\\nonumber
&=\,\frac{\tau e^2 }{3T}\int_{0}^{\infty}\frac{d\text{p}}{2\pi^2}\text{p}^2\frac{\,e^{\beta(\text{p}-\mu)}}{(1+e^{\beta(\text{p}-\mu)})^2}\,\text{E}+\,\frac{\tau e^2 }{10T}\int_{\Delta}^{\infty}\frac{d\text{p}}{2\pi^2}\frac{1}{\text{p}^2}\frac{\,e^{\beta(\text{p}-\mu)}}{(1+e^{\beta(\text{p}-\mu)})^2}\,e^2\text{B}^2\text{E}\\\nonumber
&\,\,\,+\frac{\tau e^2 }{40T^3}\int_{\Delta}^{\infty}\frac{d\text{p}}{2\pi^2}\frac{e^{4\beta(\text{p}-\mu)}-4e^{3\beta(\text{p}-\mu)}+e^{2\beta(\text{p}-\mu)}}{(1+e^{\beta(\text{p}-\mu)})^4}\,e^2\text{B}^2\text{E}\\\nonumber
&\,\,\,+\frac{\tau e^2 }{6T}\int_{\Delta}^{\infty}\frac{d\text{p}}{2\pi^2}\frac{1}{\text{p}^2}\,\frac{\,e^{\beta(\text{p}-\mu)}}{(1+e^{\beta(\text{p}-\mu)})^2}\,e^2\text{B}^2\text{E}+\frac{\tau e^2 }{12T^2}\int_{\Delta}^{\infty}\frac{d\text{p}}{2\pi^2}\frac{1}{\text{p}}\,\frac{e^{2\beta(\text{p}-\mu)}-\,e^{\beta(\text{p}-\mu)}}{(1+e^{\beta(\text{p}-\mu)})^3}\,e^2\text{B}^2\text{E}\\
=&\,\frac{\tau e^2 }{3}\left(\frac{\mu^2}{2 \pi^2}+\frac{T^2}{6}+\frac{ e^2 \text{B}^2}{16 \pi^2\mu^2}+\frac{ e^2 \text{B}^2\, T^2}{16 \mu^4}\right)\text{E}+\,\frac{\tau e^2}{3}\left(\frac{ e^2 \text{B}^2}{8 \pi^2\mu^2}+\frac{ e^2 \text{B}^2\, T^2}{8 \mu^4}\right)\text{E}.
\end{align}
Let us briefly explain the nature of the different contributions appearing above. In the first line, the first term in parentheses is due the work done on the charged particles by the electric field ($\sim \textbf{E}\cdot \textbf{v}_{\textbf{p}}$) to move them along an effective trajectory with velocity $\textbf{v}_{\textbf{p}}$. The  integrals in the second and third lines correspond to this term.  The second term in the parentheses of the first line, which corresponds to the  integrals in the fourth line, is purely originated from the anomaly ($\sim \textbf{E}\cdot \textbf{B}$). Finally in the fourth line we have split the contributions of Ohm and anomaly transport, receptively (See Appendix \ref{Long_conduc} for more details.). Let us also denote that that in writing the lower band of integrals we have considered the scheme  introduced below \eqref{T00}.

Collecting all contributions together, the longitudinal electrical conductivity $\sigma_{L}$ in the low temperature limit is given by:
\begin{equation}\label{sigma_L}
\boxed{
	\sigma_{L}=\,\frac{J^{\parallel}_{e}}{\text{E}}=\frac{e^2 \tau}{3}\left(\frac{\mu^2}{2 \pi^2}+\frac{T^2}{6}\right)+e^2 \tau\frac{\, e^2 \text{B}^2}{16 \pi^2\mu^2}\left(1+\frac{\pi^2 T^2}{\mu^2}+O(\frac{T^4}{\mu^4})\right)}
\end{equation}
The first parentheses in this formula is basically the ordinary electrical conductivity in a system of massless spin-$\frac{1}{2}$ particles in the absence of magnetic field \cite{Abrikosov}. In a kinetic system of such particles with $\epsilon(\textbf{p})=\text{p}$ and under the RTA approximation, the conductivity is given by \cite{Romatschke:2015gic}
\begin{equation}\label{sigma}
\sigma=\, \frac{\tau}{3}\,\chi=\, \frac{\tau}{3}\,\frac{\partial n}{\partial \mu}.
\end{equation} 
In the system under the consideration in this paper, the above formula can simply be evaluated via using \eqref{density}. While by neglecting the anomaly corrections in \eqref{density} we obtain exactly the first parentheses in \eqref{sigma_L}, the magnetic corrections of $\sigma_L$ cannot be found by the formula \eqref{sigma}. This simply shows that our system, when is coupled to the magnetic field, will no longer behave conformally.\footnote{We Thank N. Yamamoto for discussing on this point.} It would be interesting to investigate more on this issue in the framework of the quantum kinetic theory \cite{Hidaka:2016yjf,abbasi}.

Another important point with \eqref{sigma_L} is the positive sign of the correction term. This is sometimes referred to as the \textbf{Negative Magneto Resistivity}, NMR, (or positive magneto conductivity) \cite{Weylsemimeta:Son,Gorbar:2017lnp}. The quadratic dependence of the NMR on the magnetic field was first found in the context of chiral kinetic theory in \cite{Weylsemimeta:Son} for the Weyl semimetal. Compared to \cite{Weylsemimeta:Son}, here, we have not only considered the necessary corrections of the energy dispersion coming from the Lorentz invariance, but also we have taken into account all the sources contributing to the current, either the Ohm contribution and the anomaly one.	
Due to the generalizations were made here, the numerical factor in front of the $\text{B}^2$ differs from that of found in \cite{Weylsemimeta:Son}.
In $\sec{WSM}$, we will carefully compare the conductivities in our case to those of a Weyl semimetal.

Now let us consider a system in a uniform and steady state in the presence of a background temperature gradient.  According to the well-known Seebeck effect, if the matter in equilibrium is electrically charged, i.e. $n \ne0$, the charged particles flow from the higher temperature region to the lower one,  simply due to  the heat current driven by the temperature gradient.

As mentioned earlier,
the temperature gradient in the present case makes a role like what the electric field made in the previous subsection. So considering the second part in \eqref{delta_n_E_T} and multiplying it with $\sqrt{G}\dot{\textbf{x}}$, we compute the thermoelectric current as the following 
\begin{align}\label{j_z_th}
J^{\parallel}_{e}=
&\sum_{e}\int_{\boldsymbol{p}} \tau\,e\, \dot{\textbf{x}}_{\parallel}\,\frac{\epsilon(\text{p})-\hat{e} \mu}{T}\,\left( \frac{}{} \boldsymbol{\nabla}T\cdot \textbf{v}_{\textbf{p}}+\,e\,(\boldsymbol{\Omega_{p}}\cdot \textbf{v}_{\textbf{p}})\frac{}{}\boldsymbol{\nabla}T\cdot \textbf{B}\right)\, \frac{\partial  \tilde{n}^{(e)}_{\textbf{p}}}{\partial \text{p}}\\\nonumber
=&\,\frac{\tau e }{3T}\int_{0}^{\infty}\frac{d\text{p}}{2\pi^2}\text{p}^2(\text{p}-\mu)\frac{\,e^{\beta(\text{p}-\mu)}}{(1+e^{\beta(\text{p}-\mu)})^2}\,\zeta\,-\frac{\tau e }{8T}\int_{\Delta}^{\infty}\frac{d\text{p}}{2\pi^2}\frac{(\text{p}-\mu)}{\text{p}^2}\,\frac{\,e^{\beta(\text{p}-\mu)}}{(1+e^{\beta(\text{p}-\mu)})^2}\,e^2\text{B}^2\zeta
\end{align}
Collecting all terms in \eqref{j_z_th}, the longitudinal thermoelectric conductivity in the presence of the magnetic field reads
\begin{equation}\label{s_l}
\boxed{
	T\,\alpha_{L}=\,\frac{J^{\parallel}_{e}}{\zeta}=\frac{e\tau }{9} \, \mu T^2- e\tau\frac{ e^2  \text{B}^2 T^2}{24\mu^3}\left(1+O(\frac{T^2}{\mu^2})\right).}
\end{equation}
Again, like the conductivity formula \eqref{sigma_L}, the first term in \eqref{s_l} is related to the non-anomalous conformal matter \cite{Abrikosov}. The correction term with the negative sign is the so-called \textbf{positive magneto thermoelectric resistivity}. Similar to what happens in a Weyl semimetal   \cite{Gooth:2017mbd}, this simply shows that the anomalous effects in a Weyl fluid decrease the thermoelectric transport\footnote{Compared to the coefficient $G_T=\frac{j^{\parallel}_{e}}{\nabla T}$ defined in \cite{Gooth:2017mbd}, our thermoelectric coefficient is given by $\alpha_{1L}=-  G_{T}$.  }.

When energy is pumped into the system by the external sources, in addition to the electric current, a current of heat does flow in the system as well. The electric field participates in the flowing of the heat through the Peltier effect, while the gradient of the temperature contributes to the thermal current via ordinary thermal conduction \cite{Abrikosov}. Clearly, due to the Onsager reciprocal relations,  the coefficient of
Peltier effect is equal to that of the Seebeck effect, $s$. So in this subsection,  in addition to the coefficient of thermal conductivity, i.e. $\kappa$, we reproduce the previously found Seebeck coefficient as a check of the Onsager reciprocal relation in our system.   

In the kinetic theory, the thermal current is formally given by \cite{Abrikosov}:
\begin{equation}\label{thermal_cutrrent}
\boldsymbol{J}_{th}=\sum_{e} \int_{\textbf{p}}\sqrt{G}\, \dot{\textbf{x}}\,(\epsilon(\textbf{p})-\hat{e} \mu)\,\delta n^{(e)}_{\textbf{p}}.
\end{equation}
To evaluate it in our system, it is sufficient to multiply \eqref{delta_n_E_T} with $\sqrt{G}\dot{\textbf{x}} (\epsilon(\textbf{p})- \hat{e}\mu)$ and then perform $\sum_{e}\int_{\textbf{p}}$.
Let us start firstly by computing the thermal current induced by the external electric source. 
One writes
\begin{equation}\label{j_th_z_ohm_CME}
J^{\parallel}_{th}=-\sum_{e}\int_{\boldsymbol{p}}e \tau \dot{\textbf{x}}_{\parallel}\,(\epsilon(\textbf{p})- \hat{e}\mu)\left( \frac{}{} \textbf{E}\cdot \textbf{v}_{\textbf{p}}+\,e\,(\boldsymbol{\Omega_{p}}\cdot \textbf{v}_{\textbf{p}})\frac{}{} \textbf{E}\cdot \textbf{B}\right)\, \frac{\partial  \tilde{n}^{(e)}_{\textbf{p}}}{\partial \text{p}}
\end{equation}
which is nothing but \eqref{j_z_th} by replacing the $-\nabla T/T$ with $\text{E}$. This simply means that $\alpha_{1L}=\alpha_{2L}$ which is the manifestation of the Onsager reciprocal relation. The second contribution to the heat current comes from the thermal conduction effect. One writes
\begin{align}\label{j_th_z_th}
J^{\parallel}_{th}=&\sum_{e}\int_{\boldsymbol{p}} \tau \dot{\textbf{x}}_{\parallel}\,\frac{(\epsilon(\textbf{p})-\hat{e} \mu)^2}{T}\left( \boldsymbol{\nabla}T\cdot \textbf{v}_{\textbf{p}}+\,e\,(\boldsymbol{\Omega_{p}}\cdot \textbf{v}_{\textbf{p}})\frac{}{} \boldsymbol{\nabla}T\cdot \textbf{B}\right)\, \frac{\partial  \tilde{n}^{(e)}_{\textbf{p}}}{\partial \text{p}}\\\nonumber
=&\frac{\tau }{3T}\int_{0}^{\infty}\frac{d\text{p}}{2\pi^2}\text{p}^2(\text{p}-\mu)^2\frac{\,e^{\beta(\text{p}-\mu)}}{(1+e^{\beta(\text{p}-\mu)})^2}\,\zeta\,+\frac{\tau }{8T}\int_{\Delta}^{\infty}\frac{d\text{p}}{2\pi^2}\frac{(\text{p}-\mu)^2}{\text{p}^2}\,\frac{\,e^{\beta(\text{p}-\mu)}}{(1+e^{\beta(\text{p}-\mu)})^2}\,e^2\text{B}^2\zeta
\end{align}
So the thermal conduction coefficient in the system may be then given as the following
\begin{equation}\label{kappa_L}
\boxed{
 T\, \kappa_{L}=\,\frac{J^{\parallel}_{th}}{\zeta}=\frac{\tau (\pi T)^2}{9}\left(\frac{\mu^2}{2\pi^2}+\frac{7T^2}{10}\right)+\tau T^2\frac{\, e^2 \text{B}^2}{48\mu^2}\left(1+
	\,O(\frac{T^2}{\mu^2})\right).}
\end{equation}
As before the first part of this relation is the  thermal conductivity in the system of non-interacting massless fermions in the absence of the magnetic field \cite{Abrikosov}. The anomalous part, however, indicates that the chiral anomaly intensifies the thermal conduction in the system. In analogy with NMR, This might be called as the positive thermal conductivity. 

\subsection{Revisiting the Wiedemann-Franz law }
\label{Weidemann_Franz}
According to the Wiedemann-Franz law, at the low temperature limit, the Lorenz ratio of the thermal conductivity, $\kappa$, and the electrical conductivity, $\sigma$, is constant in a Fermi liquid \cite{Ashcroft}:
\begin{equation}
L=\, \frac{\kappa}{T \sigma}=\,\frac{\pi^2}{3 e^2}.
\end{equation}
To investigate whether the above relation holds in our present system, let us recall that the regime of applicability of the kinetic theory setup in our system was given by \eqref{regime_setup}. On the other hand, as mentioned earlier, we are interested in the low temperature regime, i.e. $T\ll \mu$, in the whole of the paper. Combining the two constraints specifies the regime of validity of our results, i.e.
$\sqrt{e\text{B}}\ll T\ll \mu$. Now to check the Wiedemann-Franz law in this regime, let us ignore about the non-anomalous parts of the conductivities and just consider the $\text{B}$ dependence of them:
\begin{eqnarray}
\kappa_{L}^{\text{B}}=\tau T^2\frac{\, e^2 \text{B}^2}{48\mu^2}\left(1+\,O(\frac{T^4}{\mu^4})\right),\,\,\,\,\,\,\,\,\,\,\,\,
\sigma_{L}^{\text{B}}=e^2 \tau\frac{\, e^2 \text{B}^2}{16 \pi^2\mu^2}\left(1+O(\frac{T^4}{\mu^4})\right)
\end{eqnarray}
Obviously
\begin{equation}
L=\, \frac{\kappa_{L}^{\text{B}}}{T^2 \sigma_{L}^{\text{B}}}=\,\frac{\pi^2}{3 e^2}.
\end{equation}
which shows that the  Wiedemann-Frans law does hold in our present regime of study for the anomalous conductivities. Physically it means that the quasi-particles in our system in the mentioned regime, scatter from the impurities in the fluid, elastically. We give more comments on this point in the summary of \sec{conclusion}.

\subsection{Side-jump and comparison with Weyl semimetals}
\label{WSM}
The negative magneto resistant in a Weyl fluid was firstly computed in \cite{Weylsemimeta:Son}  and then more accurately in \cite{Dantas:2018udo}. However in both of these works the system under the  study was a Weyl semimetal without having the Lorentz symmetry on the Lattice. Due to this reason, the authors of \cite{Weylsemimeta:Son,Dantas:2018udo} ignored the corrections of the energy dispersion given in \eqref{energy_correction} when computing the electrical conductivity.
In this subsection and under the same assumptions considered in \cite{Weylsemimeta:Son,Dantas:2018udo}, we find the other conductivity coefficients, namely $s$ and $\kappa$ (see Appendix \ref{App_WSW} for detailed computations.). Then  by comparing them with those obtained in $\sec{sec_transport}$, we discuss about the physical consequence of ignoring the corrections.

In Table.1 we have written the value of the magnetic-conductivities either with considering the corrections to the energy dispersion and without that in the context of relativistic kinetic theory. We have also given the value of the same quantities for a non-relativistic case, the Weyl semimetal case. The value of the electrical conductivity $\sigma_L$ for a Weyl semimetal, namely $\left(\frac{v}{c}\right)^3 \sigma$, has been previously found in  \cite{Dantas:2018udo}\footnote{In this Reference, the authors have scaled the Fermi velocity to unity and so what exactly they have written as the electrical conductivity is $\sigma$.}. All other coefficients are our results in the current papers. 

\begin{table}[!htb]
	\label{table one}
	\begin{center}
		\begin{tabular}{|c|c|c|c|}
			\hline
			\hline
			Longitudinal& relativistic without&& relativistic with\\
			Conductivity	&  quantum correction&WSM &   quantum  corrections \\
			& &&to second order \\
			\hline
			\hline
			& &  & \\
			Electrical: $	\sigma_{L}$&$\sigma=\frac{e^2 \tau}{3}\frac{e^2  \text{B}^2 }{5 \pi^2 \mu^2}$& $v^3 \sigma $& $\frac{15}{16} \sigma$\\
			&&& \\
			\hline
			& && \\
			Thermoelectric: $\alpha_{1L}$&$\alpha=-\frac{e \tau}{9}\frac{ 2e^2  \text{B}^2 T^2 }{ 5\mu^3}$& $v^3 \alpha$ & $\frac{15}{16}\alpha$\\
			&&& \\
			\hline
			&&  &  \\
			Thermal: 	$\kappa_{L}$	&$\kappa=\frac{\pi^2\tau}{9}\frac{ e^2  \text{B}^2 T}{ 5\mu^2}$ &$v^2 \kappa$&$\frac{15}{16}\kappa$\\
			&  & & \\
			\hline
			\hline
		\end{tabular}
	\end{center}
	\label{comparison}
	\caption{Longitudinal conductivities in the presence of magnetic field in the limit $\mu\gg T$. $v$ is the Fermi velocity in the unit of the light velocity. We are working in the relativistic system of units with $\hbar=c=1$.}
\end{table}

An interesting point with the results reported in the table is that, by considering the corrections to the energy dispersion, the value of each coefficient (given in the fourth column) turns out to be less than its counterpart which is computed with $\epsilon=p$ dispersion (given in the second column.). More interestingly, the energy corrections leads to the same decrease in the value of all three conductivities; each of them is 6.25$\%$ less than its value without energy correction.  This common behavior among all conductivities might be physically explained as it follows.

As it has been shown in \cite{Chen:2014cla}, in a system of spin-$\frac{1}{2}$ particles with definite helicity,  the Lorentz invariance implies a non-trivial modification in the Lorentz transformations. The modification is so that not only ensures the conservation of the angular momentum  in the collisions, but also implies a non-locality in the collision term in the Lorentz invariant kinetic theory, due to the side jump. Although it is always possible to find a Lorentz frame in which the side jump in  one collision does not happen \footnote{Such frame is called the no-jump frame \cite{Chen:2015gta}.}, it is hard to think about  a frame, e.g. the laboratory frame, in which the side jump does happen in none of the collisions in the system. Consequently, the decrease in the value of conductivities might be related to the side-jump effect. One may conclude that side-jump in collisions has effectively decreased the scattering time $\tau$ by 6.25$\%$.  As a result, the scatterings are happening on average in  shorter intervals than what the classical description predicts. 

In the case of Weyl semimetals, the Fermi velocity is much smaller than the velocity of light $v\ll1$; so one simply accepts that transport is weaker than that of a relativistic Weyl fluid, either without considering the relativistic corrections or with taking them into account.

\section{Ward Identities: Relations Between Conductivities}
\label{Ward}
In this section we are going to derive the set of Ward identities between one- and two-point functions in a four dimensional theory with anomalous gauge and diffeomorphism transformations.  The Ward identities in a covariant theory can be simply derived by taking the derivative of the generating functional of the theory with respect to the background gauge and metric fields. Similarly, in an anomalous theory, it is convenient to consider the desired theory as a theory living on the boundary of a one higher dimensional space time $(\mathcal{M}_5)$ within which, a topological theory, invariant under gauge and diffeomorphism transformations, lives. The covariant generating functional of such theory may be written as
\begin{equation}
W_{cov}=\,W[\partial \mathcal{M}_{5}]+\,\int_{\mathcal{M}_{5}}I^{CS}_{5}.
\end{equation}
where $I^{CS}_{5}$ is the five-form Chern-Simons associated with the gauge field, metric and a probable combination of them in five-dimensions:
\begin{equation}
I^{CS}_{5}=\,A\, \wedge\,\left[c\,F\,\wedge\,F+(1-\alpha)\frac{}{}c_m \text{tr} (R\wedge R)\right]+\,\alpha \,c_m \,F\wedge \text{tr}\left[\Gamma \wedge d\Gamma+\frac{2}{3}\, \Gamma \wedge \Gamma \wedge \Gamma\right].
\end{equation}
Here $c$ is the coefficient of triangle $U(1)^3$ anomaly  and $c_m$  is the mixed $U(1)$-gravitational anomaly coefficient. For a four dimensional system of chiral fermions,  these two coefficients are well known \cite{Jensen:2012jh}. $\alpha$ is a coefficient coming through a local gauge and diffeomorphism non-invariant contact term; it determines how the mixed anomaly is shared between $U(1)$ and the gravitational transformations and clearly does not appear in non-conservation equations of covariant energy and momentum currents (see equations \eqref{j_cov_modified}  and \eqref{T_cov_modified}  in the following).

The so-called consistent stress tensor and consistent charge current in the four dimensional theory (living on $\partial \mathcal{M}_5$) are defined by varying the generating functional $W$ with respect to the metric and gauge field on $\partial \mathcal{M}_5$, respectively
\begin{equation}\label{con_current}
J^{\mu}_{cons}=\frac{1}{\sqrt{-g}}\frac{\delta W}{\delta A_{\mu}},\,\,\,\,\,\,\,\,T^{\mu \nu}_{cons}=\frac{2}{\sqrt{-g}}\frac{\delta W}{\delta g_{\mu\nu}}.
\end{equation}
One can also define a pair of stress tensor and charge current by  varying the $W_{cov}$ (associated with $\mathcal{M}_5$) with respect to the metric and gauge field variation on $\partial \mathcal{M}_5$. To proceed let us consider an arbitrary gauge and diff transformation on the four dimensional boundary theory denoted by $\delta_{\lambda}$: 
\begin{align}\label{diff_A}
\delta _{\lambda}A_{\mu}&= \partial_{\mu}\Lambda+A_{\nu}\,\partial_{\mu}\xi^{\nu}+(\partial_{\nu}A_{\mu})\xi^{\nu}\\\label{diff_g}
\delta_{\lambda}g_{\mu\nu}&= \nabla_{\mu}\xi_{\nu}+\nabla_{\nu}\xi_{\mu}.
\end{align}
Under such transformations, the covariant generating function formally transforms as
\begin{align}\nonumber
\delta_{\lambda} W_{cov}=&\,\delta_{\lambda} W+\,\delta_{\lambda} \int_{\mathcal{M}_{5}}I^{CS}_{5}\\\label{inflow}
=&\,\delta_{\lambda} W\,+\,\int_{\partial \mathcal{M}_5} d^4x \sqrt{-g}\left\{P^{\mu}_{BZ}\, \delta_{\lambda} A_{\mu}+\frac{1}{2}P^{\mu \nu}_{BZ}\,\delta_{\lambda} g_{\mu \nu}\right\}+\,\int_{\mathcal{M}_5}(\cdots)\\\nonumber
=&\,\int_{\partial \mathcal{M}_5} d^4x \sqrt{-g}\left\{\big(J^{\mu}_{cons}+P^{\mu}_{BZ}\big) \delta_{\lambda} A_{\mu}+\frac{1}{2}\big(T^{\mu \nu}_{cons}+P^{\mu \nu}_{BZ}\big)\delta_{\lambda} g_{\mu \nu}\right\}+\,\text{inflow contribution}
\end{align}
 In above, $P^{\mu}_{BZ}$ and $P^{\mu \nu}_{BZ}$ are the Bardeen-Zumino polynomials constructed out of the Chern-Simons generating functional \cite{Bardeen}. These terms appear in the variation of the Chern-Simons five-form on the boundary.
Note that the bulk variation of the Chern-Simons term, namely the last term in the second line, vanishes in the bulk but induces an \textbf{anomaly inflow}; a flow of conserved currents from bulk to the boundary. 
The inflow contribution is given by \cite{Loganayagam:2012pz}
\begin{equation}
\int_{\mathcal{M}_5}d^5x\sqrt{-g_5}\left(\text{J}^a\delta A_{a}+\nabla_{c}\text{L}^{abc}\delta g_{ab}\right)
\end{equation}
where Latin indices run over four dimensional boundary theory coordinates as well as $\perp$, the fifth coordinate of $\mathcal{M}_5$. Let us note that while all terms in the last line of \eqref{inflow} depend explicitly on either $\partial_{\mu}\Lambda$ or $\partial_{\mu} \xi_{\nu}$, the inflow contribution includes terms just depending explicitly on the parameters $\Lambda$ and $\xi_{\mu}$, and not on their derivatives. So in order to factorize $\Lambda$ and $\xi_{\mu}$ from the integrand, just the non-inflow contributions in  \eqref{inflow} need to be integrated by part. As a result, keeping the inflow contribution aside, one may define the following so-called \textbf{covariant stress tensor} and \textbf{covariant charge current} (given in \eqref{con_current})
\begin{equation}\label{cov_current}
\begin{split}
J^{\mu}_{cov}&=\frac{1}{\sqrt{-g}}\frac{\delta W_{cov}}{\delta A_{\mu}}=\,J^{\mu}_{cons}+\,P^{\mu}_{BZ}\\
T^{\mu \nu}_{cov}&=\frac{2}{\sqrt{-g}}\frac{\delta W_{cov}}{\delta g_{\mu\nu}}=\,T^{\mu \nu}_{cons}+\,P^{\mu\nu}_{BZ}.
\end{split}
\end{equation}
These currents covariantly transform under gauge and diff transformations. 
Using the explicit expressions of the inflow contribution given in \cite{Jensen:2012jh}, the variation of the generating functional \eqref{inflow} can be simply rewritten in terms of the covrainat objects
\begin{equation}\label{delta_W_cov}
\delta_{\lambda}W_{cov}=\int_{\partial \mathcal{M}_5} d^4x \sqrt{-g}\left\{J^{\mu}_{cov}\, \delta_{\lambda} A_{\mu}+\frac{1}{2}T^{\mu \nu}_{cov}\,\delta_{\lambda} g_{\mu \nu}+\,\Lambda \,\text{J}^{\perp}+\xi_{\mu} \nabla_{\nu}\text{L}^{\perp\mu \nu}\right\}
\end{equation}
where \cite{Loganayagam:2012pz}\footnote{The superscript $\perp$ points out to the direction of the fifth dimension which is perpendicular to the field theory coordinates.}
\begin{align}
\text{J}^{\perp}=&\,\frac{1}{4} \epsilon^{\mu \nu \alpha \beta}\left(3 c F_{\mu \nu}F_{\alpha \beta}+\frac{}{}c_mR^{\lambda}_{\,\,\kappa\alpha \beta}R^{\kappa}_{\,\,\lambda\alpha \beta}\right),\\
\text{L}^{\perp\mu \nu}=&\,\frac{1}{2} c_m \, \epsilon^{\kappa \sigma \alpha \beta}F_{\kappa \sigma}R^{\mu \nu}_{\,\,\,\,\,\,\alpha \beta}.
\end{align}
Since the five dimensional theory is by construction invariant under any gauge and diff transformations, including those just acting on its four dimensional boundary theory, one can derive the anomaly equations as the following \cite{Jensen:2012jh}.
In order to obtain the Ward identities between the two-point functions we would rather working with the following two modified currents:
\begin{align}\label{density_cov_T_J}
\langle \mathcal{J}^{\mu}(x)\rangle&=\sqrt{-g}\,\langle J^{\mu}_{cov}(x)\rangle=\,\frac{\delta W[A,g]}{\delta A_{\mu}(x)}\\
\langle \mathcal{T}^{\mu \nu}(x)\rangle&=\sqrt{-g}\,\langle \mathcal{T}^{\mu\nu}_{cov}(x)\rangle=\,2\frac{\delta W[A,g]}{\delta g_{\mu\nu}(x)}.
\end{align}
After performing computations, we find that the anomaly equations can be rewritten in terms of the modified currents as the following
\begin{eqnarray}\label{j_cov_modified}
\partial_{\mu}\mathcal{J}^{\mu}&=&\,\frac{1}{4} \sqrt{-g}\,\epsilon^{\alpha \beta \rho \lambda} \left[3 c\, F_{\alpha \beta} F_{\rho \lambda}+\frac{}{}c_m\,R^{\nu}_{\,\,\kappa \alpha \beta}R^{\kappa}_{\,\,\nu \rho \lambda}\right]\\\label{T_cov_modified}
\partial_{\mu}\mathcal{T}^{\mu\nu}+\,\Gamma^{\nu}_{\mu \rho} \mathcal{J}^{\mu\rho}&=&\,F^{\mu}_{\,\,\nu}\mathcal{J}^{\nu}_{cov}+\,2 \sqrt{-g}\, c_m \nabla_{\nu} \left[\frac{1}{4}\epsilon^{\alpha \beta \rho \lambda} F_{\alpha \beta} R^{\mu \nu}_{\,\,\,\,\rho \lambda}\right]
\end{eqnarray} 
with $\Gamma^{\mu}_{\nu \rho}$ being the Christoffel symbol. 
Let us denote that $J^{\mu}$ and $T^{\mu \nu}$ are tensorial objects while  $\mathcal{J}^{\mu}$ and $\mathcal{T}^{\mu \nu}$ are densities. However, once fixing the background to be Mankowski space-time, the two descriptions coincide identically. 

The above anomaly equations are basically the Ward identities for the one-point functions $\langle J^{\mu}_{cov} \rangle$ and $\langle T^{\mu\nu}_{cov} \rangle$ which we write them for brevity  as $ J^{\mu}_{cov}$ and $ T^{\mu\nu}_{cov}$, respectively.
It is clear that the anomaly terms in the right hand side of the above two equations are coming from the anomaly inflow explained earlier.

\subsection{System coupled to the external sources}
We would like to compute the electric end heat currents in our system in the linear response regime. So we need to turn on the corresponding weak source fields, namely an electric field which induces the charge current together with a temperature gradient generating the heat current. We wish the electric field and temperature gradient vary with time as $e^{-i \omega \tau}$ in the Euclidean coordinates. To proceed, one can consider the thermodynamic state of the system in the presence of following background metric and gauge field:
\begin{align}\label{bg_metric}
ds^2&=g_{\tau\tau}(\boldsymbol{x},\tau)\,d\tau^2+\,\delta_{ij}\,dx^idx^j\\\label{bg_gauge}
\mathcal{A}&=\,\mu_E\, d\tau+  A_{i}(\boldsymbol{x},\tau)\,dx^i.
\end{align}
It is necessary for the functions $g_{\tau\tau}(\boldsymbol{x},\tau)$ and $A_i(\boldsymbol{x},\tau)$ to be such slowly varying in the space that the system is in equilibrium in every patch-wise region. The temporal component of the background gauge field, $\mu$, is the chemical potential in grand canonical ensemble. 

Considering $A_{i}(\boldsymbol{x},\tau)=\delta A_i(\boldsymbol{x})e^{- i\omega_E \tau}$ is equivalent to turning on the background electric field $E_i= - i \omega_E \delta A_i$. We assume $\omega_E \ll T_0$ with $T_0$ being the equilibrium temperature at static flat region. The condition $\omega_E \ll T_0$ ensures that $V=\partial_{\tau}$ is a Killing vector in every patch
whose size is comparable with the inverse temperature of the system $T_0^{-1}$.
The $\tau\tau$ component of the metric then induces a local temperature in the system
$T= T_0/\sqrt{g_{\tau\tau}(\boldsymbol{x})}$. Consequently, a small temperature gradient is related to a small gradient  of the $g_{\tau\tau}$ component of background metric.
Considering $T\rightarrow T+x^i\nabla_i T$ implies
\begin{equation}\label{delta_g_tt}
g_{\tau \tau}(\boldsymbol{x},\tau)=1+\delta g_{\tau\tau} =1-\,\frac{2 x^i \nabla_i T}{T}e^{-i \omega_E \tau}.
\end{equation}
At this point let us recall that the variation of the metric may be regarded as the source for the energy momentum tensor. The associated link between them is the retarded Green's function
\begin{equation}
T^{\mu \nu}\sim \,G^{\mu \nu,\alpha \beta}_{\mathcal{R}}\,\delta g_{\alpha \beta}.
\end{equation}
Following our earlier requirements, a constant temperature gradient is needed (or a constant back ground electric field) to turn on the stress tensor (or charge current) components. However, it would not be the case with the variation of the $\tau\tau$ component given in \eqref{delta_g_tt}. In order to remove the $\boldsymbol{x}$-dependence of $g_{\tau\tau}(\boldsymbol{x})$, one may demand an appropriate diffeomorphism  act on \eqref{bg_metric} and \eqref{bg_gauge}. It is readily shown that the transformations \eqref{diff_A} and \eqref{diff_g} with 
\begin{equation}
\xi_{\mu}=\left(i \,                                                                                                                                                                                                                                                                                                                                                                                                                                                                                                                                                                                                                                                                                                                                                                                                                                                                                                                                                                                                                                                                                                                                                                                                                                                                                                                                                                                                                                                                                                                                                                                                                                                                                                                                                                                                                                                                                                                       x^i \nabla_i T/\omega_E T,\boldsymbol{0}\right)e^{-i \omega_E \tau},\,\,\,\Lambda=0
\end{equation}
give rise to the following changes in the background metric and gauge field (denoted by $\delta'$)
\begin{align}
\delta' g_{\tau\tau}&=\frac{2 x^i \nabla_i T}{T}e^{-i \omega_E \tau},\,\,\,\,\,\delta'g_{\tau i}=\,-\frac{  \nabla_i T}{i \omega_E T}e^{-i \omega_E \tau}\\
\delta'A_{\tau}&=\mu_E\,\frac{ x^i \nabla_i T}{T}e^{-i \omega_E \tau},\,\,\,\,\,\delta'A_i=\,-\mu_E\,\frac{ \nabla_i T}{i \omega_E T}e^{-i \omega_E \tau}.
\end{align}
Obviously, $\delta g_{\tau\tau}+\delta' g_{\tau\tau}=\,0$. So the variation of the generating functional (given by \eqref{delta_W_cov}) simplifies to
\begin{equation}
\delta W_{cov}=\int d^3 x d\tau \sqrt{-g_E}\left\{(T_{cov}^{\tau i}+\mu_E J_{cov}^i)\frac{-\nabla_i T}{i \omega_E T}-
J_{cov}^{i}\frac{E_i}{i \omega_E}\right\}.
\end{equation}
By replacing $\tau=i t$, $\omega_{E}=-i \omega$,   $\mu_E=-i \mu$ and $E_i \rightarrow -i E_i$ we can go back to the Minkowski space-time
\begin{equation}\label{heat_from_W}
\delta W_{cov}=-i\int d^3 x dt \sqrt{-g}\left\{(T_{cov}^{t i}-\mu J_{cov}^i)\frac{-\nabla_i T}{i \omega T}+J_{cov}^{i}\frac{E_i}{i \omega}\right\}.
\end{equation}
It should be
noted that we have dropped the terms explicitly depending on $\boldsymbol{x}$ from the integrand, since they do not contribute to the integral. One can also simply show that for the same reason, $\xi_{\mu} \nabla_{\nu}\text{L}^{\perp\mu \nu}$ hes been ignored to be written in the integrand. \footnote{The time dependent factor $e^{-i \omega \tau}$ has been absorbed in the background fields and the imaginary time has been transformed back to the real one.  }

The above computation has an important outcome. It is well-known that in a non-anomalous system, the heat and electric currents coupled to the background temperature gradient and the electric field are $Q^i=T^{ti}- \mu J^{i}$ and $J^{i}$, respectively \cite{Hartnoll:2009sz,Herzog:2009xv}.  We have shown that in an anomalous system, the same expression can be used for the heat and electric current, however,  we have to be careful to write such currents for the covariant stress tensor and covariant current.

\subsection{Conductivities and Ward identities}
Since we consider the theory at a finite chemical potential, the finite charge density  then mixes the heat and electric currents. So the Ohm's law must be generalized to
\begin{equation}\label{generalized_Ohm}
\begin{pmatrix}
J_i  \\
Q_i\end{pmatrix}=\begin{pmatrix}
\sigma_{ij} &T \alpha_{ij} \\
T \alpha_{ij} &T \kappa_{ij} 
\end{pmatrix}\,\begin{pmatrix}
E_j  \\
-\nabla_j T/T\end{pmatrix}.
\end{equation}              
From now on we omit the subscript "$cov$" from $J^{\mu}_{cov}$ and $T^{\mu \nu}_{cov}$ and simply refer to them as $J^{\mu}$ and $T^{\mu \nu}$.
Inserting $E_{i}=- i \omega (\delta A_i+ \mu \delta  g_{ti})$ and $-\nabla_i T/T= i \omega \delta g_{ti}$ in \eqref{generalized_Ohm} and considering \footnote{Here by $\delta A_i$ and $\delta g_{ti}$, we mean the total variation of $A_i$ and $g_{ti}$, including both $\delta$ and  $\delta'$ variations mentioned in previous subsection.}
\begin{equation}\label{two_point}
G^{J_iJ_j}_\mathcal{R}=\,- \frac{\delta \mathcal{J}_{i}}{\delta A_j},\,\,\,\,\,\,\,\,\,\,\,\,G^{Q_iJ_j}_\mathcal{R}=\,- \frac{\delta \mathcal{Q}_{i}}{\delta A_j},\,\,\,\,\,\,\,\,\,\,\,\,\,G^{Q_iQ_j}_\mathcal{R}=\,- \frac{\delta \mathcal{Q}_{i}}{\delta g_{tj}}
\end{equation}
with $\mathcal{J}^i=\sqrt{-g} J^i$ and $\mathcal{Q}^i=\sqrt{-g}Q^i$, one simply reads the conductivities as  
\begin{equation}\label{conductivity_Kubo}
\sigma_{ij}({\omega})=\frac{e^2\,G^{J_iJ_j}_\mathcal{R}(\omega)}{i \omega},\,\,\,\,\,\,\,\,\,\,\,\,\,\,\,\,\,\,\,\,T \alpha_{ij}({\omega})=\frac{e\,G^{Q_iJ_j}_\mathcal{R}(\omega)}{i \omega},\,\,\,\,\,\,\,\,\,\,\,\,\,\,\,\,T \kappa_{ij}({\omega})=\frac{G^{Q_iQ_j}_\mathcal{R}(\omega)}{i \omega}.
\end{equation}
In the following we will show that the longitudinal conductivities $\sigma_{33}\equiv\sigma_L$, $\alpha_{33}\equiv\alpha_L$ and $\kappa_{33}\equiv\kappa_L$ are not fully independent;  once specifying one of them, the Ward identities ensure that the other two are immediately specified. 

Since we are interested in relations between the longitudinal conductivities, we assume the system to be in the presence of a constant magnetic field $B$ directed in the 3-direction and find Ward identities for  the two-point functions. Before proceeding, let us recall that in the equilibrium of our system, the only non-vanishing components of the energy-momentum and charge one-point functions are as the following \footnote{The anomalous transport coefficients $\sigma_{B}$ and $\sigma_{B}^{\epsilon}$ will be introduced in the next section.}
\begin{equation}
\langle \mathcal{J}^{0}\rangle=n,\,\,\,\,\,\langle \mathcal{J}^{3}\rangle=\sigma_{B}B,\,\,\,\,\,\langle \mathcal{T}^{00}\rangle=\epsilon+p,\,\,\,\,\,\langle \mathcal{T}^{03}\rangle=\sigma_{B}^{\epsilon}B.
\end{equation}
In order to evaluate \eqref{two_point}, let us couple the system to background fields $\delta A_z$ and $\delta g_{tz}$. Varying equation \eqref{T_cov_modified} with respect to $\delta A_z$ and $\delta g_{tz}$ we find the two following identities at $k=0$
\begin{align}\label{condition_1}
G_{\mathcal{R}}^{T_{03}J_{3}}+\langle \mathcal{J}^0\rangle=&\,0\\ \label{condition_2}
G_{\mathcal{R}}^{T_{03}T_{03}}+\langle \mathcal{T}^{00}\rangle=&\,0.
\end{align}
Now we are able to show how the coefficients $\alpha_{L}$ and $\kappa_{L}$ depend on $\sigma_{L}$. From the Kubo formulas \eqref{conductivity_Kubo} and by using \eqref{condition_1}
and \eqref{condition_2} we arrive at
\begin{align}\label{alpha_zz}
T\,\alpha_{L}&=\frac{e}{i \omega}\left\{G^{T_{03}J_{z}}_{\mathcal{R}}-e\mu\frac{}{}G_{\mathcal{R}}^{J_3J_3}\right\}=\,-\frac{e\,n}{i \omega}-\,\frac{\mu}{e}\frac{}{}\, \sigma_{L}\\\label{kappa_zz}
T\,\kappa_{L}&=\frac{1}{i \omega}\left\{G^{T_{03}T_{03}}_{\mathcal{R}}-2\mu G^{T_{03}J_{3}}_{\mathcal{R}}+\mu^2\frac{}{}G_{\mathcal{R}}^{J_3J_3}\right\}=\,-\frac{\epsilon+p -2\mu n}{i \omega}+\,\frac{\mu^2}{e^2}\, \sigma_{L}.
\end{align}
In \cite{Hartnoll:2009sz,Herzog:2009xv}, similar relations were found in a $2+1$ dimensional non-anomalous system. We have however shown that in an anomalous system while two new equilibrium currents are induced, namely $\mathcal{J}^0$ and $\mathcal{T}^{03}$, the relation between longitudinal conductivities will not get change compared to the non-anomalous case studied in \cite{Hartnoll:2009sz,Herzog:2009xv}.

\subsection{Ward identities and consistency with the chiral kinetic theory results}
Let us recall that in \sec{kinetic} we computed the longitudinal conductivities in a system of right-handed Weyl fermions coupled to an external weak magnetic field.  On the other hand, the constraint relations obtained in the previous subsection have to be satisfied by the longitudinal conductivities in every arbitrary anomalous system, including the system of free massless fermions earlier studied in the current paper. To investigate the latter, let us start by considering the right hand side of  \eqref{alpha_zz}.
Using \eqref{density} and \eqref{sigma_L} and replacing $i/\omega$ with $\tau$, we may write
\begin{align}\nonumber
\tau\, e\, n-\frac{\mu}{e}\,\sigma_{L}&=\tau \,e\, \left(\frac{\mu^3}{6 \pi ^2}+\frac{\mu T^2}{6 }\right)+\tau \,e\,\frac{e^2\text{B}^2}{16 \pi^2 \mu}\left(1+\frac{\pi^2T^2}{3\mu^2}+\,O(\frac{T^4}{\mu^4})\right)\\\nonumber
&\,\,\,-\frac{\mu}{e}\frac{e^2 \tau}{3}\left(\frac{\mu^2}{2 \pi^2}+\frac{T^2}{6}\right)-\frac{\mu}{e}e^2 \tau\frac{\, e^2 \text{B}^2}{16 \pi^2\mu^2}\left(1+\frac{\pi^2 T^2}{\mu^2}+O(\frac{T^4}{\mu^4})\right)\\
&=\frac{e\tau }{9} \, \mu T^2- e\tau\frac{ e^2  \text{B}^2 T^2}{24\mu^3}\left(1+O(\frac{T^2}{\mu^2})\right)\equiv\,T \alpha_{L}
\end{align} 
which coincides with \eqref{s_l}.
Analogously, one can show that \eqref{kappa_zz} holds for the results obtained in \sec{kinetic}. The consistency between our kinetic results with the Ward identities in the limit $k\rightarrow 0$ simply shows that the computations performed in \sec{kinetic} are all valid in the hydrodynamic limit by replacing the relaxation time parameter $\tau$ with $i/\omega$. 

One central point in \sec{kinetic} was that the quantum corrections had to be taken into account to second order to observe the phenomena like the negative magneto-resistivity. In the above we saw that such corrections are indeed important for the conductivities to obey the Ward identities as well. This means that the conductivities found in Weyl semimetals without quatum corrections, like what computed in \cite{Son:2012zy,Dantas:2018udo,kim} and developed in Appendix \ref{App_WSW}, do not satisfy the constraint relations \eqref{alpha_zz} and \eqref{kappa_zz}.

\subsection{Comparison with Lucas et al \cite{Lucas:2016omy}}
In \cite{Lucas:2016omy}, a "covariant" theory of thermoelectric transport in weakly disordered Weyl semimetals has been presented.  Their
hydrodynamic theory consists of relativistic fluids at each Weyl node which are coupled together by small inter-valley scattering, and long-range Coulomb interactions. They enter the dissipation  via adding the relaxation terms to the right hand side of the hydrodynamic equations. The mentioned terms characterize the rate of the intervalley transfer of charge, energy
and momentum due to relative imbalances of the temperature or chemical potential between the nodes.   While their conductivities contain quadratic contributions of the magnetic field, the authors ignore the second order corrections of hydrodynamics. It is apparently due to the regime of parameters assumed in that paper \footnote{We thank R. Davison for pointing this out to us.}. 

Demanding physical requirements in a simple case of a Weyl semimetal with 2 valley fluids, the authors of \cite{Lucas:2016omy} reduce all the unknown coefficients in their model just to three ones. The latter can be analytically computed in a weakly intercating Weyl gas with weak intervalley scattering; a case similar to what studied in the current paper.\footnote{In fact, like $\tau$ in our kinetic computations, the time scale set by the relaxation coefficients of \cite{Lucas:2016omy} is the longest one in the problem.} Finally,
in the limit $\mu \gg T$, the leading order magneto-conductivities are reported to satisfy the following relations in \cite{Lucas:2016omy}
\begin{align}\label{Sachdev_alpha}
\alpha_{L}&=\,\frac{\pi^2 T}{3e}\,\frac{\partial \sigma_{L}(\mu, T=0)}{\partial \mu}\\\label{sachdev_kappa}
\kappa_{L}&=\frac{\pi^2 T}{3e^2}\,\sigma_{L}
\end{align}
At this point we would like to investigate whether the magneto-conductivities found in our paper satisfy the above relations.
From eq.\eqref{sigma_L}, the longitudinal electrical conductivity may at $T=0$ is written as
\begin{equation}\label{}
\sigma_{L}(\mu,T=0)=\frac{e^2 \tau}{3}\frac{\mu^2}{2 \pi^2}+e^2 \tau\frac{\, e^2 \text{B}^2}{16 \pi^2\mu^2}
\end{equation}
whose derivative with respect to $\mu$ (multiplies with $\pi^2/3e$) is given by
\begin{equation}\nonumber
\frac{\pi^2 T}{3e}\,\frac{\partial \sigma_{L}(\mu, T=0)}{\partial \mu}=\, \frac{e\tau }{9} \, \mu T- e\tau\frac{ e^2  \text{B}^2 T}{24\mu^3}\equiv \alpha_L\,\, \text{found in} \,\,\eqref{s_l}.
\end{equation} 
This simply shows that our results obey the equation \eqref{Sachdev_alpha}. The second relation, namely \eqref{sachdev_kappa}, was already verified when we was studying the validity of Wiedemann-Franz law in \sec{Weidemann_Franz}.

Now we may be tempted to conclude that the constraints \eqref{Sachdev_alpha} and \eqref{sachdev_kappa} are related to the general constraints  \eqref{alpha_zz} and \eqref{kappa_zz}, obtained from Ward identities. Let us recall that the latter were obtained for a general system in the presence of anomalies. To show the existence  of such relation, by comparing \eqref{alpha_zz} with \eqref{Sachdev_alpha} in the system of free massless fermions, we interestingly find $\sigma_L$ just by knowing the expression of the charge density in equilibrium. To proceed, let us equate \eqref{alpha_zz} with \eqref{Sachdev_alpha} at $T=0$ by considering \eqref{density} and replacing $-i \omega$ with $1/\tau$. We find the following differential equation
\begin{equation}\label{diff_equ}
e\left(\frac{\mu^3}{6 \pi ^2 }+\frac{e^2\text{B}^2}{16 \pi^2 \mu}\right)\tau-\,\frac{\mu}{e} \sigma_L(\mu,T=0)=\,\frac{\pi^2 T^2 }{3e}\,\frac{\partial \sigma_{L}(\mu, T=0)}{\partial \mu}
\end{equation}
Solving this equation we find in general
\begin{equation}
\sigma_{L}^{sol}=\,e^2\frac{\mu^2\tau}{6\pi^2}\left(1-\frac{2\pi^2T^2  }{3\mu^2}\right)+e^{-\frac{3\mu^2}{2\pi^2T^2  }}\left(C \,+e^2 \tau\frac{3e^2 \text{B}^2}{32 \pi^4 T^2}\text{ExpIntegralEi}\big(\frac{3\mu^2}{2\pi^2T^2  }\big)\right)
\end{equation}
with $C$ being a constant. Since the equation \eqref{diff_equ} is valid in the limit $T\rightarrow 0$, we take the same limit from the above solution 
\begin{equation}
\sigma_{L}(\mu, T=0)=\,e^2 \tau\frac{\mu^2}{6\pi^2}+\,e^2 \tau\frac{e^2\text{B}^2}{16 \pi^2 \mu^2}.
\end{equation}
This is obviously nothing but the expression \eqref{sigma_L} at $T=0$.

In summary, in this subsection we showed that our results about magneto-transport in a Weyl fluid are consistent with the hydrodynamic model of \cite{Lucas:2016omy} in the same limit. That the comparison between the constraint equation of \cite{Lucas:2016omy} in the special system of free Weyl fermions, i.e. \eqref{Sachdev_alpha}, with our general ones, i.e. \eqref{alpha_zz}, gives precisely the previously found $\sigma_{L}$ in \eqref{sigma_L} shows the importance of the general relations found in previous subsection. In other words, once the charge density of the fermionic system is given, we can find the longitudinal conductivity just by considering the Ward identities together with the constraint equations of \cite{Lucas:2016omy}.

\section{Summary, Conclusion and Outlook}
\label{conclusion}
Let us firstly review what we found in this paper. The main idea for starting this work was to study the magneto-transport in a "relativistic" Weyl fluid in the framework of (chiral) kinetic theory. Compared to the analogous study in a non-relativistic Weyl semimetal \cite{Weylsemimeta:Son}, we needed to compute the appropriate quantum corrections to the dispersion of Weyl particles in the phase space. Since the magneto-conductivities were expected to quadratically depend on the magnetic field, we were to find the second order correction to the energy dispersion as well. Doing so, in \sec{central_result} we arrived at our central result in \eqref{central}. Let us denote that such corrections were originally coming from the Berry flux of a Berry monopole located at the origin of the momentum space.

Then we argued that such correction would affect on the thermodynamic quantities of the system. To show the latter rigorously, we computed the energy-momentum tensor and the charge current components \eqref{T_00}, \eqref{pressure}, \eqref{T_11} and \eqref{density}. The main problem which we encountered with throughout the computations, was the emergence of some IR divergences in the phase space integrals. To overcome this, we divided each integral into two parts; 1. a  no $\text{B}$-dependent part  and 2. a magnetic part. While the first part had nothing to do with the Berry monopole, the second part was actually a quantum mechanical correction caused by its Berry flux. Then in accordance with the chiral kinetic theory requirements, we gave a scheme to regulate the divergences and compute the integrals.

Our scheme is simply that to the first part of the integrals, namely the no-$\text{B}$ dependent part discussed in previous paragraph,  all the states in the momentum space contribute (see the lower bound of the integral \eqref{I_1} for instance.). For computing the second part of integrals, however, our scheme  is that, not only it would be needed to exclude the states with momenta less than  $\sim \sqrt{e\text{B}}$ \cite{Stephanov:2012ki}, but it would be necessary to get an IR cut-off $\Delta_{\text{B}}$ in the momentum space so that
$\sqrt{e\text{B}}\ll \Delta_{\text{B}}$.
 Only states with $\Delta_{\text{B}}\lesssim\text{p}$ contribute to the second part of the integrals (see the lower bound of the integrals in \eqref{I_2} and \eqref{I_3}.). It can be also seen that this scheme works truly when $\text{B}\rightarrow 0$.

 As an application of the thermodynamic quantities obtained by the above scheme, specifically the enthalpy density, we then computed the magneto-electrical conductivity in the system. Our computation was based on the fact that in the presence of the magnetic field, due to the chiral magnetic effect (CME) current, the density of chiral charges could not remain conserved. We computed the heat density produced by the CME current during the annihilation of the chiral charges. Then by relating the produced heat to the enthalpy density \eqref{enthalpy_CKT}, we read the magneto electrical conductivity \eqref{prediction}. 
 
 The main part of our computations about the magneto-transport has been done in \sec{sec_transport}. In this section by use of the linear response theory and under the relaxation time approximation we computed not only the electrical conductivity \eqref{sigma_L}, but also the thermo-electric \eqref{s_l} and the thermal \eqref{kappa_L} conductivities as well. All of the computations were done analytically in the limit $\mu \gg T$. Interestingly, the magnetic part of the electrical conductivity \eqref{sigma_L} was turned out to be in complete agreement with the one obtained from the enthalpy density in \eqref{prediction}.
 
   Our results show that in the limit $\mu \gg T$ the Wiedemann-Franz law identically holds\footnote{In \cite{kim} the general structure of the magneto-conductivities have been found in Weyl metal, although, no explicit result has been reported. With no considering the quantum correctins It has been also argued that the Wiedemann-Franz law breaks down in general in the Weyl metal.}. Consequently the relativistic Weyl fluid at low temperature limit behaves like a Fermi liquid. In fact the main reason behind this behavior is the special type of the relaxation time approximation we assumed  in the system.
In the language of the Weyl semimetals, this corresponds to weak intervalley scattering in system of weakly interacting Weyl fermions \cite{Lucas:2016omy}.

By repeating the computations of \sec{sec_transport} for a system of Weyl fermions without energy corrections, namely by taking $\epsilon(\textbf{p})=\text{p}$, we arrived at an interesting result. Let us refer to such conductivities by a superscript "nc" denoting that they are non-corrected. We found that (see \sec{App_WSW} and table \ref{comparison})
\begin{equation*}
\frac{\sigma_{L}}{\sigma_{L}^{nc}}=\frac{\alpha_{L}}{\alpha_{L}^{nc}}=\frac{\kappa_{L}}{\kappa_{L}^{nc}}=\,\frac{15}{16}.
\end{equation*} 
This observation is in agreement with the side-jump picture. Due to the side-jump in scatterings of particles, the scattering time $\tau$ decreases on average. Since  the side-jump comes from the quantum corrections, one expects the scattering time and consequently the conductivity decrease 

 Quantitatively, the above common ratio, $15/16$, for all the conductivities suggests that they might be linearly dependent to each other. To confirm the idea we found anomalous Ward  identities at infinite long wave length limit and thereby, obtained the expected linear constraint relations between conductivities \eqref{alpha_zz} and \eqref{kappa_zz}. To our knowledge, such relations had not been obtained for a 3+1 dimensional anomalous system before.

As a first consistency check, we showed the conductivities found in \eqref{sigma_L}, \eqref{s_l} and \eqref{kappa_L} obey the constraint relations \eqref{alpha_zz} and \eqref{kappa_zz}. This confirms the necessity of the second order correction of the energy dispersion as well as the scheme we developed to regulate the integrals in the kinetic theory. We also checked that the conductivities computed without the quantum corrections, namely those corresponded to a non-relativistic WSM, would not obey the constraints mentioned above (see \sec{App_WSW}). 

As another consistency check, we compared our results with those of the "covariant" model of \cite{Lucas:2016omy}. Their system is a WSM that becomes the same as our system if considered with weakly intervalley scattering. In this limit, the general magneto-conductivities obtained from their model, are constrained by two relations. These relations differ from the constraints obtained from Ward identities in our paper. 
We showed that once the charge density in equilibrium is found \eqref{density}, the comparison between two latter sets of relations gives rise to a differential equation for the electrical conductivity. Its solution at $\mu\gg T$ is exactly the expression for the electrical conductivity we obtained previously in \eqref{sigma_L} and \eqref{prediction}.

In summary, while we computed the magneto-conductivities, $\sigma_{L}$, $\alpha_{L}$ and $\kappa_{L}$, via the linear response method, the electrical conductivity $\sigma_{L}$ was computed, additionally, via two another approaches as well. First, from the enthalpy density of the equilibrium and second, from the comparison of the constraints of Ward identities with the relations of \cite{Lucas:2016omy}.

In addition to its necessity for studying the chiral transport, the second order correction found in this paper might be important for further developments of chiral kinetic theory. Recently the chiral kinetic theory of Weyl fermions has been derived from quantum filed theory in \cite{Hidaka:2016yjf}. In the mentioned paper, the first order quantum correction of the energy of Weyl particles has been found via finding the following modified on-shell condition for them \footnote{For the sake of concreteness, we have restored the factor $\hbar$ and its powers.}
\begin{equation}
\textbf{p}^2-\epsilon(\textbf{p})^2+\hbar \, e\textbf{B}\cdot \hat{\textbf{p}}=\,O(\hbar^2\,e^2\text{B}^2)
\end{equation}
Interestingly, when $\epsilon(\textbf{p})$ in this formula is replaced with the second order corrected one, namely \eqref{energy_correction_second}, the equality does still hold! It means that the right hand side of the above equation vanishes at least to $O(\hbar^3\,e^3\text{B}^3)$. Two questions arise immediately;
first, does it mean that no side-jump term is needed in the perturbative solution of Wigner function at second order in \cite{Hidaka:2016yjf}? Second,  how about the higher orders? Does the above equation hold to all orders in quantum corrections? If yes, could that be related to non-renormalization of chiral anomaly \cite{abbasi2}?   Answering to each of these questions may help to better understanding of the relation between chiral kinetic theory and quantum field theory anomalies. 

As discussed around \eqref{Boltzaman_hydro}, the regime of study in this paper is nothing other than the hydrodynamics. Let us recall that in the standard hydrodynamic derivative expansion, the magnetic field is counted as a one derivative object. So in order to study the magneto-transport in the universal framework of the hydrodynamics,  \footnote{In \cite{Landsteiner:2014vua} such study has been done, however, just for the case of electrical conductivity and just by considering the first order hydrodynamics. See also   \cite{Roychowdhury:2015jha,Sun:2016gpy,Rogatko:2018lhn,Rogatko:2018moa} for similar studies.} it is needed to keep the derivatives to second order in the constitutive relations. \footnote{The first and second order hydrodynamic corrections and constraints on their corresponding transport coefficients have been widely studied in the literature\cite{Son:2009tf,Kharzeev:2011ds,Kovtun:2012rj,Landsteiner:2016led,Bhattacharya:2011tra,Neiman:2010zi,Buzzegoli:2018wpy,Buzzegoli:2017cqy,Hernandez:2017mch,Sadooghi:2016ljd}.} 
Once having found the magneto-conductivities from the second order hydrodynamics, then one can simply apply them to well-known physical systems. An interesting example in the weak coupling regime is the system of free fermions studied in the current paper. At strong coupling, there are certain holographic systems dual to an anomalous system in the presence of the magnetic field \cite{DHoker:2009ixq}. \footnote{The thermo-electric transport in strong coupling has been studied in a lot of papers including \cite{Davison:2016ngz,Donos:2014cya,Li:2018ufq,Mokhtari:2017vyz}.} We leave more investigation on the magneto-transport from hydrodynamics to a future work \cite{abbasi}.

Finally, it would be also interesting to repeat the computations of the current paper in a more realistic case in which the Weyl fluid contains two types of chiral fermions with opposite chiralities at different chemical potentials.  This might be more relevant to quark gluon plasma physics in the heavy ion collision experiments.

\section*{Acknowledgment}
We would like to thank M. Chernodub, R. Dantas, R. Davison,  D. Kharzeev, K. Landsteiner, S. Pu and  N. Yamamoto for useful discussions. We would also like to thank S. Ben-Abbas,  A. Ghazi,  P. Janghorban, K. Kiaei, K. Naderi, P. Rezaei and A. Shahrabi for discussion. We are grateful to A. Davody for reading the manuscript.
N. A. would like to thank Prof. F. Ardalan for his encouragements. 
N. A. would also like to thank M. Alishahiaha and M. Mohammadi for their supports. 
 O. Tavakol would like to thank school of particles and accelerators of IPM for hosting during this work. N. A. would like to thank  University of Tours for hospitality when some part of the current work was being done.
\appendix

\section{Regularizing the integrals}
\label{regular}
We first split the integral of energy density to the three parts.
\begin{align}\label{T_00_App}\nonumber
T^{00}
=&\,\int_{\Delta_{\text{B}}}^{+\infty}\frac{d\text{p}}{2 \pi^2}\frac{1}{1+e^{\beta(\text{p}-\mu)}}\left[\text{p}^3-\frac{\text{B}^2}{8 \text{p}}+\frac{\text{B}^2}{24 T^2}\frac{e^{\beta(\text{p}-\mu)}(T+\text{p})+e^{2\beta(\text{p}-\mu)}(T-\text{p})}{(1+e^{\beta(\text{p}-\mu)})^2}\right]\\
=&\,I_1+I_2+I_3
\end{align}
The first part is in fact the non-magnetic contribution to the energy in the region between two spheres in \fig{fig_sphere}. In the $\mu\gg T$ limit the anti-particle states do not contribute and we find 
\begin{align}\label{I_1}
I_1=\int_{0}^{+\infty}\frac{d\text{p}}{2\pi^2}\frac{\text{p}^3}{1+e^{\beta(\text{p}-\mu)}}
=\,-\frac{3 T^4}{\pi^2}\text{PolyLog}[4,e^{-\beta \mu}]
=\,\frac{\mu^4}{8\pi^2}+\frac{\mu^2 T^2}{4}+\frac{7\pi^2T^4}{120}+O(T^4e^{-\beta \mu}).
\end{align}
 The second part in \eqref{T_00_App} is the divergent one and needs more explanation.
To perform this integral, we first make an integration by part to change the distribution function to a symmetric function around $\text{p}=\mu$. We may write
\begin{align}\nonumber
I_2=&\,-\frac{\text{B}^2}{16\pi^2}\left(\frac{\log\text{p}}{1+e^{\beta(\text{p}-e\mu)}}\right)\bigg|_{\Delta_{\text{B}}}^{+\infty}-\,\frac{\text{B}^2}{8T}\int_{\Delta_{\text{B}}}^{+\infty}\frac{d\text{p}}{2\pi^2}\log\text{p}\,\frac{e^{\beta (\text{p}-e\mu)}}{(1+e^{\beta (\text{p}-e\mu)})^2}\\\nonumber
=&\,\,\frac{\text{B}^2}{16\pi^2}\,\log\Delta_{\text{B}}-\,\frac{\text{B}^2}{8T}\int_{-\infty}^{+\infty}\frac{Tdx}{2 \pi^2}\log(\mu+T x)\,\frac{e^x}{(1+e^x)^2}\,+O(\text{B}^2e^{-\beta \mu})\\\nonumber
=&\,\,\frac{\text{B}^2}{16\pi^2}\,\log \Delta_{\text{B}}-\,\frac{\text{B}^2}{8}\int_{-\infty}^{+\infty}\frac{dx}{2 \pi^2}\left(\log \mu+\frac{T x}{\mu}-\frac{T^2x^2}{2\mu^2}+\frac{T^3 x^3}{3\mu^3}-\frac{T^4 x^4}{\mu^4}+\cdots\right)\,\frac{e^x}{(1+e^x)^2}\\\label{I_2}
=&\,\,-\frac{\text{B}^2}{16\pi^2}\left(\log\frac{\mu}{\Delta_{\text{B}}}-\frac{\pi^2}{6}\frac{T^2}{\mu^2}-\frac{7\pi^4}{60}\frac{T^4}{\mu^4}+\,O(\frac{T^5}{\mu^5})\right)
\end{align}
In the second line we have changed the integrand variable as $x=\beta(\text{p}-e\mu)$. It is obvious that in the limit $\mu\gg T$, the lower bound of $x$ goes to $-\infty$. Similar to what often used in condensed matter physics, we have exploited the Sommerfeld expansion \cite{Ashcroft} and expanded $\log(\mu+Tx)$ in powers of $x$ in the integrand (third line above).  

Finally we compute the third part of \eqref{T_00_App} as the following
\begin{align}\nonumber
I_3=&\frac{\text{B}^2}{24 T^2}\int_{\Delta_{\text{B}}}^{+\infty}\frac{d\text{p}}{2 \pi^2}\frac{T\,e^{\beta (\text{p}-\mu)}}{(1+e^{\beta(\text{p}-\mu)})^2}+\frac{\text{B}^2}{24 T^2}\int_{\Delta_{\text{B}}}^{+\infty}\frac{d\text{p}}{2 \pi^2}\frac{\text{p}\,(e^{\beta (\text{p}-\mu)}-e^{2\beta (\text{p}-\mu)})}{(1+e^{\beta(\text{p}-\mu)})^3}\\\nonumber
=&\,\,\frac{\text{B}^2}{24 T^2}\int_{0}^{+\infty}\frac{d\text{p}}{2 \pi^2}\frac{T\,e^{\beta (\text{p}-\mu)}}{(1+e^{\beta(\text{p}-\mu)})^2}+\frac{\text{B}^2}{24 T^2}\int_{0}^{+\infty}\frac{d\text{p}}{2 \pi^2}\frac{\text{p}\,(e^{\beta (\text{p}-\mu)}-e^{2\beta (\text{p}-\mu)})}{(1+e^{\beta(\text{p}-\mu)})^3}+O(\text{B}^2e^{-\beta \mu})\\\nonumber
=&\,\,\frac{\text{B}^2}{24 T^2}\int_{-\infty}^{+\infty}\frac{Tdx}{2 \pi^2}\frac{T\,e^{x}}{(1+e^{x})^2}+\frac{\text{B}^2}{24 T^2}\int_{-\infty}^{+\infty}\frac{Tdx}{2 \pi^2}\frac{(\mu+Tx)\,(e^{x}-e^{2x})}{(1+e^{x})^3}+O(\text{B}^2e^{-\beta \mu})\\\label{I_3}
=&\,\,\frac{\text{B}^2}{24 \pi^2}\,+O(\text{B}^2e^{-\beta \mu})
\end{align}
Collecting \eqref{I_1}, \eqref{I_2} and \eqref{I_3}, $T^{00}$ turns out to be as given in \eqref{T_00}. 

It should be noted that the expressions of pressure \eqref{pressure}  , $T^{11}$ and $T^{22}$ in \eqref{T_11_22_33} and also the charge density \eqref{density}  have all been found through the above procedure in this paper.

In summary, what forced us to treat with $I_2$ differently in comparison with $I_1$ and $I_3$ is that the Sommerfeld expansion just works well whenever the phase space integral takes the following form
\begin{equation}\label{general_integral}
\int_{\Delta_{\text{B}}}^{+\infty}\frac{d\text{p}}{2\pi^2}\,g(\text{p})\,\frac{e^{n \beta(\text{p}-\mu)}}{(1+e^{ \beta (\text{p}-\mu)})^m},\,\,\,\,\,\,m>n>0.
\end{equation}
In the limit $\mu\gg T$, by changing the variable as $x=(\text{p}-\mu)/T$, the above integral can be written as
\begin{equation}
\int_{-\infty}^{+\infty}\frac{T dx}{2\pi^2}\bigg(g(\mu)+ T x\, g'(\mu)+\frac{T^2x^2}{2!}g''(\mu)+\cdots\bigg)\frac{e^{n x}}{(1+e^x)^m}
\end{equation} 
and can be performed analytically.

\section{Longitudinal conductivities in Weyl fluid}
\label{Long_conduc}
In this subsection we will compute the longitudinal electrical conductivity in detail. The main steps of the computations of the other conductivities would be then the same. 
As it can be seen, all the integrals  in \eqref{j_z_ohm_CME} are in the form of \eqref{general_integral}. So by considering $x=(\text{p}-\mu)/T$, we may write
 \begin{align}\nonumber
J^{\parallel}_{e}&=\,\frac{\tau e^2}{3T}\int_{-\infty}^{+\infty}\frac{T dx}{2\pi^2}(T^2 x^2 +\mu^2)\,\frac{e^x}{(1+e^x)^2}\text{E}+\,\frac{\tau e^2}{10T}\int_{-\infty}^{+\infty}\frac{T dx}{2\pi^2}\frac{1}{\mu^2}\left(1+\frac{3T^2 x^2}{\mu^2}\right)\,\frac{e^x}{(1+e^x)^2}e^2\text{B}^2\text{E}\\\nonumber
&\,\,\,\,+\frac{\tau e^2}{40T^3}\int_{-\infty}^{+\infty}\frac{Tdx}{2\pi^2}\frac{e^{4x}-4e^{3x}+e^{2x}}{(1+e^x)^4}\, e^2\text{B}^2\text{E}\\\nonumber
&\,\,\,\,+\frac{\tau e^2 }{6T}\int_{-\infty}^{\infty}\frac{T dx}{2\pi^2}\frac{1}{\mu^2}\,\left(1+\frac{3T^2 x^2}{\mu^2}\right)\,\frac{e^x}{(1+e^x)^2}e^2\text{B}^2\text{E}\\
&\,\,\,\,+\frac{\tau e^2 }{12T^2}\int_{-\infty}^{\infty}\frac{Tdx}{2\pi^2}\frac{1}{\mu}\left(-\frac{2Tx}{\mu}-\frac{4T^3x^3}{\mu^3}\right)\,\frac{e^{2x}-\,e^{x}}{(1+e^{x})^3}\,e^2\text{B}^2\text{E}
 \end{align}
where we have kept contributing terms to the power of four in the expansion over $1/\mu$. After evaluating the integrals, one arrives at the last line of \eqref{j_z_ohm_CME}.

It is worth-mentioning that while in the above computation we find the result with corrections that correct $\sigma_{L}(\text{B}=0)$ to the order of $\text{B}^2T^2/\mu^6$, for the other conductivities like 
\eqref{s_l} and \eqref{kappa_L}, we keep just terms that correct the non-magnetic part to the order of $\text{B}^2/\mu^4$. The reason for this is that we need to have the expressions to the orders which consistently satisfy the constraints \eqref{alpha_zz} and \eqref{kappa_zz}.
\section{Conductivities in WSM}
\label{App_WSW}
In the absence of the quantum corrections to the energy dispersion ($\epsilon(\textbf{p})=\text{p}$), the electric current in \eqref{j_z_ohm_CME} turns out to be as the following  
\begin{equation}\label{j_z_ohm_CME_Weyl}
\begin{split}
\sigma_{L}^{WSM}=&\,\frac{\tau e^2}{3T}\int_{\Delta_{\text{B}}}^{+\infty}\frac{d\text{p}}{2\pi^2}\text{p}^2\,\frac{e^{\beta (\text{p}-\mu)}}{(1+e^{\beta (\text{p}-\mu)})^2}+\tau e^2\, \frac{2e^2\text{B}^2}{15T}\int_{\Delta_{\text{B}}}^{+\infty}\frac{d\text{p}}{2\pi^2}\frac{1}{\text{p}^2}\,\frac{e^{\beta (\text{p}-\mu)}}{(1+e^{\beta (\text{p}-\mu)})^2}\\
=&\,\frac{\tau e^2}{3T}\int_{-\infty}^{+\infty}\frac{T dx}{2\pi^2}(T x +\mu)^2\,\frac{e^x}{(1+e^x)^2}\\
&\,\,\,\,\,\,\,\,+\,\tau e^2 \frac{2e^2\text{B}^2}{15T}\int_{-\infty}^{+\infty}\frac{Tdx}{2\pi^2}\frac{1}{\mu^2}\left(1-\frac{2T x}{\mu}+\frac{3T^2 x^2}{\mu^2}+\cdots\right)\,\frac{e^{x}}{(1+e^{x})^2}\\
=&\frac{\tau e^2 }{3}\left(\frac{\mu^2}{2 \pi^2}+\frac{T^2}{6}\right)+\tau e^2 \frac{\, e^2 \text{B}^2}{15
	\pi^2\mu^2}\left(1+O(\frac{T^2}{\mu^2})\right)
\end{split}
\end{equation}
Again as before the contribution of the cut-off is negligible. So the lower bound of integrals goes to zero.
In performing the second integral above in the limit $\mu\gg T$, it was only sufficient to get the leading term in the Sommerfeld expansion. This is equivalent with the following replacement
\begin{equation}
\mu\gg T:\,\,\,\,\,\,\,\,\,\frac{e^{\beta (\text{p}-\mu)}}{(1+e^{\beta (\text{p}-\mu)})^2}\,\,\rightarrow\,\,\delta(\text{p}-\mu).
\end{equation}
Similarly the thermoelectric coefficient reads
\begin{equation}\label{j_z_th_Weyl}
\begin{split}
T\,\alpha_{L}^{WSM}=&\,\frac{e\tau}{3T^2}\int_{\Delta_{\text{B}}}^{+\infty}\frac{d\text{p}}{2\pi^2}\text{p}^2(\text{p}-\mu)\,\frac{e^{\beta (\text{p}-\mu)}}{(1+e^{\beta (\text{p}-\mu)})^2}+\,e \tau\frac{2e^2\text{B}^2}{15T^2}\int_{\Delta_{\text{B}}}^{+\infty}\frac{d\text{p}}{2\pi^2}\frac{\text{p}-\mu}{\text{p}^2}\,\frac{e^{\beta (\text{p}-\mu)}}{(1+e^{\beta (\text{p}-\mu)})^2}\\
&=\frac{e\tau }{9}\mu T^2+e \tau\frac{2\, e^2 \text{B}^2T^2}{45
	\pi^2\mu^3}\left(1+O(\frac{T^2}{\mu^2})\right)
\end{split}
\end{equation}
and finally the thermal conductivity reads
\begin{equation}\label{j_th_z_th_Weyl}
\begin{split}
T\,\kappa_{L}^{WSM}=&\,\frac{\tau}{3T^2}\int_{\Delta_{\text{B}}}^{+\infty}\frac{d\text{p}}{2\pi^2}\text{p}^2(\text{p}-\mu)^2\,\frac{e^{\beta (\text{p}-\mu)}}{(1+e^{\beta (\text{p}-\mu)})^2}+\,
 \tau\frac{2e^2\text{B}^2}{15T^2}\int_{\Delta_{\text{B}}}^{+\infty}\frac{d\text{p}}{2\pi^2}\frac{(\text{p}-\mu)^2}{\text{p}^2}\,\frac{e^{\beta (\text{p}-\mu)}}{(1+e^{\beta (\text{p}-\mu)})^2}\\
&=\frac{\tau (\pi T)^2}{9}\left(\frac{\mu^2}{2\pi^2}+\frac{7T^2}{10}\right)+ \tau T^2\frac{\, e^2 \text{B}^2}{45
	\pi^2\mu^2}\left(1+O(\frac{T^2}{\mu^2})\right).
\end{split}
\end{equation}
At this point, it is worth-mentioning that the above conductivities do not satisfy the constraints obtained from the Ward identities, namely \eqref{alpha_zz} and \eqref{kappa_zz}.
This shows the importance of the above computations, since the three different conductivities are independent in this case. Although the Onsager reciprocity does still hold and Seebeck and Peltier coefficients coincide identically.

\bibliographystyle{utphys}
\providecommand{\href}[2]{#2}\begingroup\raggedright\endgroup

\end{document}